\documentclass[aps,prl,10pt,longbibliography,twocolumn,showpacs,amsmath,amssymb,superscriptaddress,floatfix,nobibnotes]{revtex4-2}
\usepackage[utf8]{inputenc}
\usepackage{graphicx}  
\usepackage{soul}
\usepackage{bm}       
\usepackage{amsfonts,amsmath,amssymb,mathtools}  
\usepackage[colorlinks=true,breaklinks=true,allcolors=blue]{hyperref}
\usepackage{orcidlink}
\usepackage{physics}
\usepackage{dsfont}
\usepackage{enumitem}

\begin{document}

\title{Shortcuts to adiabaticity with a quantum control field}

\author{Emma C. King \orcidlink{0000-0002-6696-3235}}
\affiliation{Theoretische  Physik,  Universit\"at  des  Saarlandes,  D-66123  Saarbr\"ucken,  Germany}
\author{Giovanna Morigi \orcidlink{0000-0002-1946-3684}}
\affiliation{Theoretische  Physik,  Universit\"at  des  Saarlandes,  D-66123  Saarbr\"ucken,  Germany}
\affiliation{Center for Quantum Technologies (QuTe), Saarland University, Campus, 66123 Saarbr\"ucken, Germany}
\author{Rapha\"el Menu \orcidlink{0000-0001-7641-9922}}
\affiliation{Theoretische  Physik,  Universit\"at  des  Saarlandes,  D-66123  Saarbr\"ucken,  Germany}
\affiliation{CESQ/ISIS (UMR 7006), CNRS and Universit\'{e} de Strasbourg, 67000 Strasbourg, France}

\date{\today}

\begin{abstract}
Quantum adiabatic dynamics is the crucial element of adiabatic quantum computing and quantum annealing. Shortcuts to adiabaticity enable acceleration of the computational time by suppressing unwanted non-adiabatic processes with designed classical fields. Here, we consider quantum state transfer in the Landau-Zener model, which exemplifies the key elements of quantum adiabatic dynamics. We argue that non-adiabatic transitions can be suppressed by autonomous quantum dynamics, which involves coupling the Landau-Zener qubit to a second quantum system. By tuning the coupling strength, the composite quantum dynamics can reduce the probability of unwanted processes by more than two orders of magnitude. This is a prime example of control where the quantum properties of the control fields are key for implementing shortcuts to adiabaticity.
\end{abstract}

\maketitle

\noindent \textit{Introduction}.---The Landau-Zener (LZ) model is a paradigm of a time-dependent quantum mechanical process that is exactly solvable and encompasses the fundamental principles of quantum adiabatic transfer \cite{Landau, Zener, Stueckelberg, Majorana_1932}. For these reasons, the LZ model provides key insights into quantum thermodynamic principles \cite{Barra:2016}, quantum phase transitions \cite{Damski_2005,de_grandi_adiabatic_2010}, and quantum annealing dynamics \cite{Santoro_2002,Dziarmaga:2010,Albash_2018}, serving as the workhorse for identifying strategies for accelerating adiabatic transfers~\cite{Damski_2005} and unveiling the mechanisms that determine their ultimate speed limits~\cite{Mullen_1989,Hegerfeldt:2013}. Some of these strategies utilize classical control fields, which induce processes that suppress adiabatic transitions, thereby implementing so-called ``shortcut to adiabaticity" protocols~\cite{Chen:2010,GueryOdelin_2019,turyansky_2025}. This can be enforced through nonlinearities \cite{Tian:2016,Li:2018,
Pick:2021,Sveistrys_2025,Deffner:2025}, feedback \cite{Ding:2025} and reinforcement learning \cite{Ding:2021, Ai:2022}. Perfect cancellation of the non-adiabatic transitions is achieved by counterdiabatic protocols \cite{Demirplak:2008,Berry_2009,del_Campo_2013,Opatrny_2014,Giannelli:2014,Kolodrubetz_2017}, among which the control protocol is known for the LZ dynamics \cite{Berry_2009}. Approximated versions thereof can be realized by means of Floquet engineering \cite{Claeys:2019,Petiziol:2024} and strategies based on optimal control \cite{Demirplak:2008,Cepaite:2023}.

Thermal reservoirs have also been shown to facilitate adiabatic transfers \cite{Shimshoni:1991,Shimshoni_1993,Amin_2008,de_Vega_2010}. This led to the identification of open-quantum-system approaches to shortcuts to adiabaticity \cite{Avron_2010,Avron_2011,Arceci_2017,Dupays:2021,King_2023} and to the formulation of strategies for reservoir-induced transitionless quantum driving \cite{Vacanti_2014,Alipour:2020}. In these protocols the incoherent dynamics is engineered through the coupling with a second quantum system, which can possess the features of a reservoir \cite{Tamascelli:2017,Ciccarello_2022} or undergo dissipative dynamics itself \cite{Poyatos_1996,Horak:1997,Menu_2022,King_2023}. The majority of these models disregard quantum correlations between the LZ qubit and the second quantum system, treating the latter as memoryless. Nevertheless, there are indications that memory effects can lead to more favorable speed limits \cite{Wild_2016,Alipour:2020,Menu_2022,Butler:2024}. In particular, it has been posited that entanglement between the LZ qubit and the {\it quantum} control field during the dynamics can be resourceful for shortcuts to adiabaticity \cite{Menu_2022}. This hypothesis is corroborated by studies of the multilevel LZ model \cite{Arceci_2017,Ivakhnenko:2023,Stehli_2023} and on quantum annealing protocols designed by auxiliary systems \cite{Pino_2020,Cai_2024}.

\begin{figure}[b]
    \centering
    \includegraphics[width=0.9\linewidth]{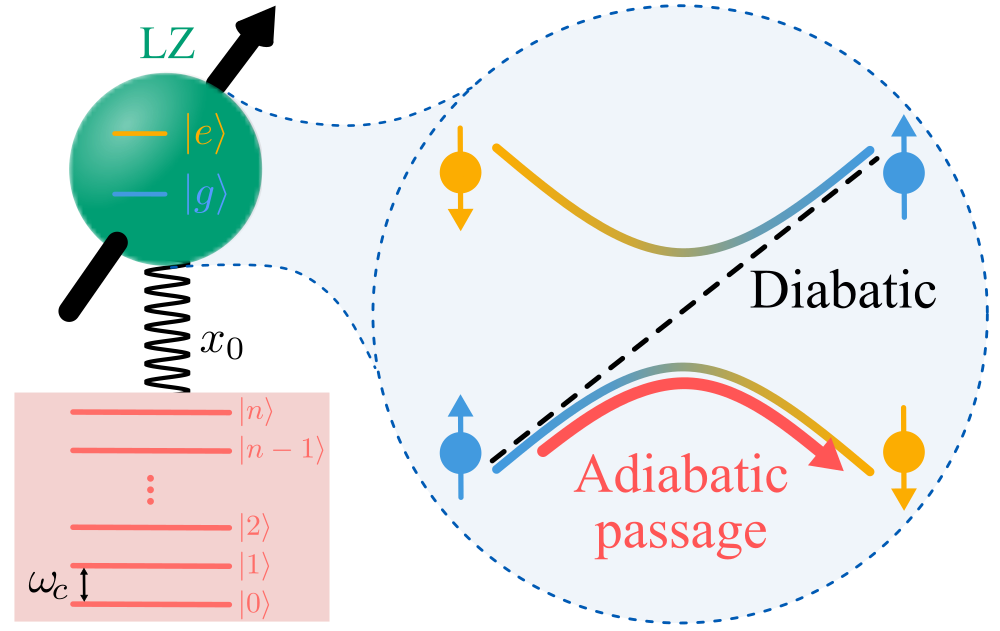}
    \caption{Interference-assisted shortcut to adiabaticity: The ultra-strong coupling of a LZ qubit to a quantum field (bosonic or fermionic) leads to high-fidelity adiabatic passage, even for fast sweeps where the isolated LZ dynamics would be in the diabatic regime. Here, $x_0$ scales the strength of the coupling, and $\omega_c$ is the eigenfrequency of the quantum field.}
    \label{fig:1}
\end{figure}

In this work, we analyze the fidelity of quantum state transfer in the LZ model. We assume that the LZ qubit couples linearly to a second quantum system, which we interchangeably denote by \textit{quantum field} or \textit{spectator}.  We demonstrate that, when the LZ qubit dynamics is deep in the diabatic regime, where quantum state transfer is inefficient, the composite dynamics of the LZ qubit and spectator can correct for the diabatic processes and reduce the probability that they occur by more than two orders of magnitude. The protocol is illustrated in Fig.~\ref{fig:1}. High-fidelity transfer is realized when the coupling between qubit and spectator is in the so-called ultra-strong coupling regime \cite{Frisk:2019,Ridolfo_2021}. The protocol does not require the time-dependent control of the quantum field. Instead, it is optimized by choosing the field eigenfrequency and the coupling strength with the LZ qubit, and is robust against fluctuations in the physical parameters.
\begin{figure*}
    \centering
    \includegraphics[width=\textwidth]{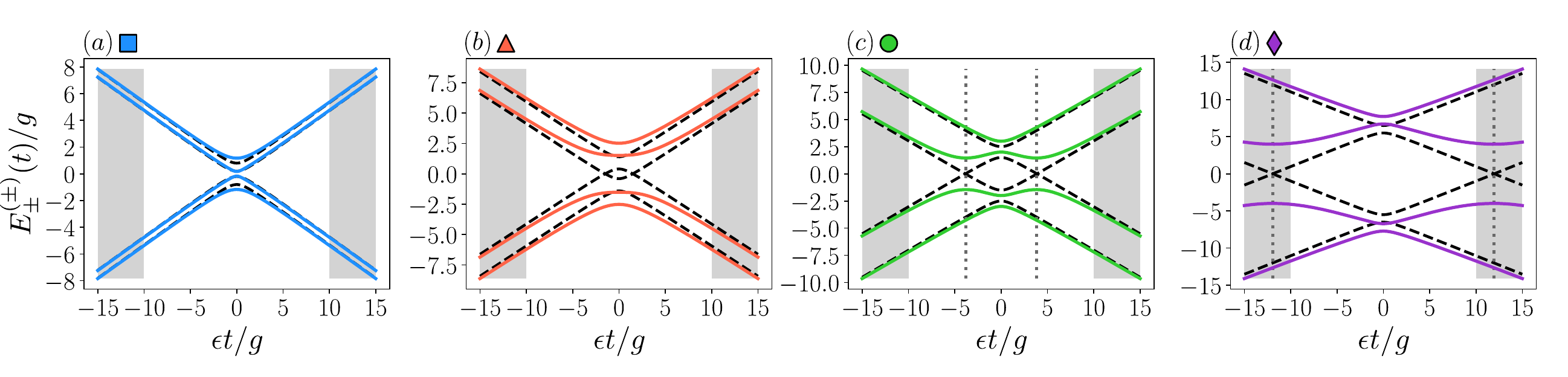}
    \caption{Instantaneous energy branches $E_\pm^{(\pm)}(t)$ of the qubit-field Hamiltonian $\hat{H}$ of Eq.\,\eqref{eq:H} (solid lines) for exemplary values of $x_0$ and $\omega_c$ and protocol time window $t_f=10g/\epsilon$. Times outside this window, $t>t_f$, are shaded gray. Dashed lines are the qubit-field system instantaneous eigenenergies at $x_0=0$. Subfigures exemplify Regime (I), where the coupling with the field tends to close the minimal gap (subplot (a), $x_0=\omega_c=0.6g$); Regime (II), where the coupling with the field tends to promote high-fidelity transfer (Subplots (b) and (c)). In (b) the coupling increases the gap at $t=0$ and flattens the spectrum in its vicinity ($x_0=\omega_c=1.8g$). In (c) ``side gaps'' (indicated by gray dotted lines) occur at $ t_\mathrm{sg}\approx \pm 3.8g/\epsilon$ ($x_0=1.5g$, $\omega_c=4g$). Subplot (d) illustrates Regime (III), where the anticrossings at $t_\mathrm{sg}\approx 12g/\epsilon>t_f$ occur outside the protocol time window ($x_0=4g$, $\omega_c=12g$). The symbols on top help to locate the corresponding spectrum in the phase diagram of Fig.~\ref{fig:4}.}
    \label{fig:2}
\end{figure*}

\noindent\textit{The model}.---The quantum state $|\Psi\rangle_t$ of LZ qubit and field at time $t$ is determined by the Schr\"odinger equation, governed by the Hamiltonian acting on the composite Hilbert space of qubit ($s$) and field ($f$): 
\begin{equation}
\label{eq:H}
\hat H=\hat{H}_{s}(t)\otimes \hat{\mathbb{I}}_f+\hat H_{\rm int}+\hat{\mathbb{I}}_s\otimes \hat H_{f}\,,    
\end{equation} 
where $\hat{\mathbb{I}}_i$ is the identity operator in the corresponding Hilbert space. $\hat{H}_s(t)$ is the LZ Hamiltonian
\begin{equation}
    \hat{H}_s(t)=\left[\epsilon t \,\hat{\sigma}^z + g \,\hat{\sigma}^x \right]/2\,,\label{eq:1}
\end{equation}
where $\hbar=1$ and $\hat{\sigma}^j$ are the Pauli matrices ($j=x,y,z$). Hamiltonian \eqref{eq:1} describes the dynamics of a spin $1/2$ (qubit) in a classical magnetic field with a constant component along $x$ and a time-dependent amplitude (sweep) along $z$, which varies at a constant rate~$\epsilon$. The sweep is of finite duration $t\in[t_i,t_f]$ with a symmetric crossing, $t_f=-t_i$ \cite{Vitanov_Garraway1996_LandauZenerModelEffects}. The instantaneous Hamiltonian $\hat H_s(t)$ is diagonal in the eigenbasis $\vert \pm \rangle_t$ with eigenenergies $E_\pm (t) = \pm \frac{1}{2}\sqrt{(\epsilon t)^2 + g^2}$. In the absence of the interaction with the spectator, the probability that the LZ qubit, initially prepared in the lower branch, undergoes a transition to the upper branch is $\mathcal{I}_{\mathrm{LZ}}(t_f,t_i)=|_{t_f}\langle +|\psi\rangle_{t_f}|^2$, with $|\psi\rangle_t$ the qubit quantum state at time~$t$. For $t_i\rightarrow-\infty$ and $t_f\to\infty$, the probability (infidelity) takes the compact form $\mathcal{I}_{\mathrm{LZ}}=\exp\left(-\pi g^2/2\epsilon\right)$, showing that adiabatic transfer is warranted for \mbox{$g^2/\epsilon\gg 1$} \cite{Vitanov1999_TransitionTimesLandauZener,Vitanov_Garraway1996_LandauZenerModelEffects}. Evidently, $\mathcal{I}_{\mathrm{LZ}}$ decreases monotonously by increasing the gap $g$ and/or decreasing the sweep rate $\epsilon$ \cite{Ivakhnenko:2023}. 

In what follows, we consider the diabatic regime, in which $g^2/\epsilon<1$ and $\mathcal{I}_{\mathrm{LZ}}>0.2$. For convenience, we now assume that the spectator field is a second qubit at the transition frequency $\omega_c$, with $\hat H_f=\omega_c\hat{\tau}^z/2$, and $\hat\tau^j$ denoting the Pauli matrices acting on the states of the spectator Hilbert space. The qubit-spectator interaction is the minimal coupling Hamiltonian 
\begin{equation}
    \hat{H}_\mathrm{int}=x_0\,\hat{\sigma}^x\otimes \hat{\tau}^x\label{Eq:Hint},
\end{equation}
and is scaled by the parameter $x_0$ with the dimension of an energy. 

Hamiltonian \eqref{eq:H}, and variations thereof, have been discussed in several contexts. A trivial but instructive case is found for $\epsilon=0$: the interaction Hamiltonian \eqref{Eq:Hint} commutes with the qubit Hamiltonian \eqref{eq:1}, and the spectator field undergoes a dynamics conditioned by the state of the qubit. Measurement of the state of the spectator then provides information on the state of the qubit. This is the basic principle underpinning quantum non-demolition measurements \cite{Braginsky:1980, haroche2013exploring, YUKALOV2012253}. This concept has been extended in Refs.~\cite{Menu_2022,King_2023} for designing adiabatic protocols in which the instantaneous interaction commutes with $\hat H_s(t)$ at all times. For $\epsilon\neq 0$, when the spectator's eigenfrequency vanishes ($\omega_c=0$), the field is stationary in the eigenbasis of $\hat{\tau}^x$. By judiciously selecting the state of the spectator, the qubit undergoes a LZ dynamics with the effective gap $g'=g+x_0$ at the anticrossing $t=0$. This results in a rescaling of the adiabaticity parameter ${g'}^2/\epsilon$, which increases the fidelity of the transfer (see also \cite{Sveistrys_2025} for a similar case). In its complete form, for $\omega_c\neq 0$, the dynamics governed by Eq.\,\eqref{eq:H} spans the four-dimensional Hilbert space of qubit and spectator. We will show that this is key for promoting high-fidelity quantum state transfer in the diabatic regime of the LZ model, $g^2/\epsilon<1$. 

\noindent{\it Adiabatic basis and spectrum}.---Relevant frequency scales of the coupled dynamics of Eq.\ \eqref{eq:H} can be extracted in the adiabatic basis of the instantaneous eigenstates of $\hat H(t)$.  This can be done analytically, as we show in the supplemental material (SM) \cite{SMref}.
Figure \ref{fig:2} displays the eigenenergies as a function of $t$ for different values of $x_0$ and $\omega_c$. In general, the two branches $E_{\pm}(t)$ of the LZ model are doubled. The coupling $x_0$ entangles the qubit and spectator states. At fixed $x_0$, tuning $\omega_c$ can give rise to a single anticrossing at $t=0$ or to multiple anticrossings at $t=0$ and $|t|> 0$. 
 
Two limits are immediate: for $|t|\to\infty$ the qubit and spectator decouple and $\hat H(t)$ is diagonal in the eigenbasis of $\hat\sigma^z\otimes\hat\tau^z$, while at $t=0$ the qubit Hamiltonian \eqref{eq:1} reduces to $\hat H_s(0)=g\hat\sigma^x/2$. In the instantaneous eigenbasis of $\hat H_s(0)$, such that $\hat\sigma^x\vert\pm\rangle_0=\pm\vert\pm\rangle_0$, the global Hamiltonian reads
$\hat H(0)=\sum_{s=\pm1}\vert s\rangle_0\langle s\vert\otimes\hat h_f^{(s)}$, with $\hat h_f^{(s)}=sg\hat{\mathbb{I}}_f/2 + s x_0\hat\tau^x +\omega_c\hat\tau^z/2$ defined on the Hilbert space of the spectator qubit. The eigenenergies take the simple form $E_{\pm}^{(s)}(0)=(s g\pm\Delta_0)/2$, with
\begin{equation}\label{eq:gap}
		\Delta_0=\sqrt{4x_0^2+\omega_c^2}\,,
\end{equation}
showing that the splitting at $t=0$ is set by the LZ minimal gap $g$ and the spectator splitting $\Delta_0$. In general, the spectrum features an anticrossing at $t=0$. The value $|\omega_c|= \sqrt{g\Delta_0}$ separates the case in which there is either a single anticrossing at $t=0$ or two additional ones at $t_{\mathrm{sg}}=\pm(\omega_c^2-g^2\Delta_0^2/\omega_c^2)^{1/2}/\epsilon$ (see SM \cite{SMref}). 

Figure \ref{fig:2} illustrates the possible spectral forms, obtained by varying only the spectator qubit's frequency $\omega_c$ and the coupling strength $x_0$. Figure~\ref{fig:2}(a) shows the spectrum for relatively weak coupling, regime (I), where upper and lower branches of the unperturbed Hamiltonian hybridize at the center and the resulting minimal gap $\sim|g-\Delta_0|$ is smaller than the bare LZ gap $g$. This behavior occurs for $0<\Delta_0<\Delta_c^{(1)}$, where for $\Delta_c^{(1)}= 2g$ the gap equals $g$. Since the minimal gap decreases by increasing $x_0$, we expect that the coupling degrades the transfer efficiency. Figures~\ref{fig:2}(b) and~\ref{fig:2}(c,d) depict two spectra at $\Delta_0>\Delta_c^{(1)}$ showing different behaviors about the region $t=0$, determined by the ratio $|\omega_c|/\sqrt{g\Delta_0}$.
For $|\omega_c|\lesssim \sqrt{g\Delta_0}$, Fig.~\ref{fig:2}(b), the coupling enlarges the splitting at $t=0$ between upper and lower branches. This is regime (II) and we expect that this behavior will tend to suppress diabatic transitions. For $|\omega_c|\gtrsim \sqrt{g\Delta_0}$, Fig.~\ref{fig:2}(c,d), the spectrum now features two distinct gaps at $t_{\rm sg}$, whose location is controlled by the sweep rate $\epsilon$. The transfer efficiency will generally depend on whether the annealing window contains the side gaps. Since we are interested in fast transfer, i.e.\ a small annealing window, we focus on the regime of Fig.~\ref{fig:2}(d), where the side gaps lie outside the annealing window and the two central branches tend to close away from $t=0$. We denote this regime by (III) and expect that the size of the side gaps will limit the efficiency of the protocol. This leads to a second bound $\Delta_c^{(2)}(\epsilon)$, such that for $\Delta_0>\Delta_c^{(2)}$ the efficiency of the protocol will start to deteriorate. In the SM we show that when the gaps are inside the annealing window, the spectral properties of Fig.~\ref{fig:2}(c) indeed lead to high-fidelity transfer. 

This discussion indicates that high-fidelity transfer could be realized for short annealing times if one tunes the parameters $x_0,\,\omega_c$ such that the dynamics is in regime (II). This requires $\Delta_0<\Delta_c^{(2)}$ and simultaneously $\Delta_0>\Delta_c^{(1)}$, leading to the necessary condition $\Delta_c^{(2)}(\epsilon)>\Delta_c^{(1)}$. Interestingly, inequality $\Delta_c^{(2)}(\epsilon)>\Delta_c^{(1)}$ is satisfied for $g<\sqrt{\epsilon}$, therefore, deep in the diabatic regime of the bare LZ model. Moreover, the parameter window $\Delta_0\in[\Delta_c^{(1)},\Delta_c^{(2)}]$ grows with $\epsilon$ \cite{SM1}. Somehow counterintuitively, our analysis suggests that, in this regime, the efficiency of the transfer increases with the sweep rate $\epsilon$.

\begin{figure}[b]
    \centering
    \includegraphics[width=\columnwidth]{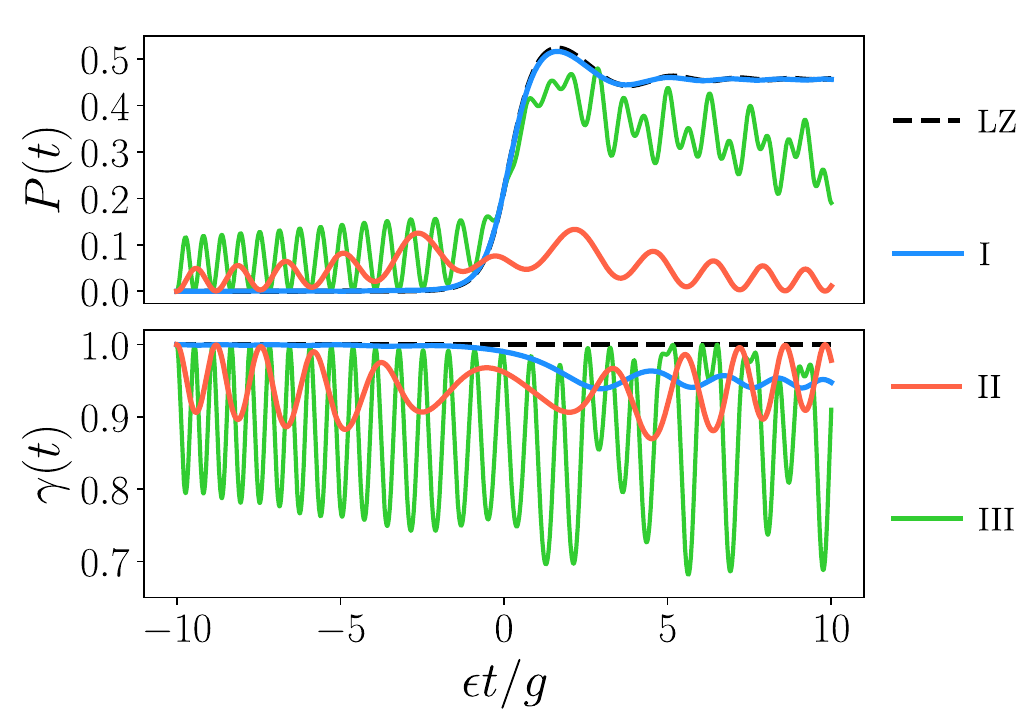}
    \caption{Evolution of the transition probability $P(t)$, Eq.\,\eqref{eq:transition_prob}, and purity $\gamma(t)$, Eq.\,\eqref{eq:purity}, for the qubit-field dynamics governed by Hamiltonian \eqref{eq:H} and in the diabatic regime, $\epsilon = 2g^2$. Three different choices of $x_0$ and $\omega_c$ are shown, illustrating the regimes of the annealing dynamics: (I) $x_0 = 0.1g$ and $\omega_c = x_0$, (II) $x_0 = 1.2g$ and $\omega_c = 0.6x_0$, (III) $x_0 = 4g$ and $\omega_c = 3x_0$. The black dashed line represents the bare LZ dynamics ($x_0=0$). The qubit-field system is initialized at time $t_i = -10g/\epsilon$ in state $\vert\Psi\rangle_0 = \vert -\rangle_{t_i}\otimes\vert\downarrow\rangle $ and evolved to time $t_f = -t_i$. Extending the time window damps the oscillations and further reduces $P(t)$ in regimes (II) and (III); see SM \cite{SMref}.}
    \label{fig:3}
\end{figure}

\noindent{\it Dynamics}.---The quantum state of the qubit is generally described by the density matrix $\hat{\rho}(t)={\rm Tr}_f\{|\Psi\rangle_t\langle\Psi|\}$ obtained by tracing out the degrees of freedom of the spectator. We consistently initialize in the product state $\vert\Psi\rangle_0=\vert-\rangle_{t_i}\otimes\vert\downarrow\rangle$, with $\vert\downarrow\rangle$ denoting the spectator's ground state. To assess the efficiency of the transfer protocol, we determine the probability $P(t)$ that the qubit is in the excited branch $\vert + \rangle_t$ of the bare LZ Hamiltonian, which we interchangeably denote by infidelity:
\begin{equation}\label{eq:transition_prob}
    P(t)= \mathrm{Tr}\{\hat\rho(t)|+\rangle_t\langle +|\}\,.
\end{equation}
The probability $P(t)$ refers solely to the excitation of the LZ model, irrespective of the state of the quantum field. Entanglement between qubit and field is characterized by means of the purity
\begin{equation}\label{eq:purity}
    \gamma(t) = \mathrm{Tr}\{\hat{\rho}^2(t)\}\,,
\end{equation}
which is linked to the R\'enyi entropy $S_2 = -\ln \gamma$ and is a witness of entanglement \cite{Adesso_2012}. The maxima $\gamma=1$ signal that the qubit is in a pure state. 

\begin{figure}[t]
    \centering
    \includegraphics[width=\columnwidth]{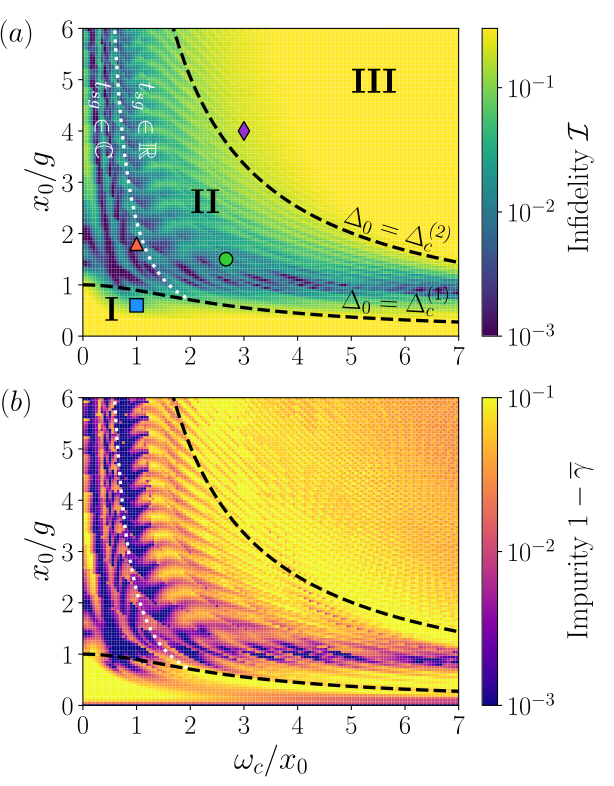}
    \caption{Protocol's efficiency: (a) Infidelity $\mathcal{I}=P(t_f)$ \eqref{eq:transition_prob} and (b) impurity $1-\overline{\gamma}=1-\gamma(t_f)$ \eqref{eq:purity} of the final state as a function of the parameters $x_0/g$ and $\omega_c/x_0$ for $t_f\approx 10g/\epsilon$. Annealing procedure and choice of parameters are as in Fig.~\ref{fig:3}. The dashed lines indicate the critical values $\Delta_c^{(1)}= 2g$ and $\Delta_c^{(2)}$, separating the three regimes. The dotted curve splits regime (II) into two subregions: Below the curve there is only a single anticrossing at $t=0$. Above the curve new anticrossings emerge at $t=\pm|t_\mathrm{sg}|$. Markers in (a) refer to the corresponding spectra in Fig.~\ref{fig:2}. For the chosen annealing time the infidelity oscillates at the end of the protocol. The values we report are obtained by taking the local minimum of $P(t_f)$ and $\gamma(t_f)$ in the window $t_f\in[8g/\epsilon,10g/\epsilon]$. The results are consistent with other optimization procedures, such as taking the mean of $P(t)$ and $\gamma(t)$ over the final 10\% of the annealing window. They are also insensitive to the spectator's initial state. See SM~\cite{SMref} for further details.}
    \label{fig:4}
\end{figure}

Figure \ref{fig:3} displays the evolution of $P(t)$ in the diabatic regime of the LZ model (here $\epsilon=2g^2$). For this choice, the bare LZ dynamics predicts that $P(t)$ exceeds 0.45 at the end of the transfer. The system dynamics in regime~(I) remain effectively unchanged under coupling to the spectator qubit. In regime~(III) we observe a slow improvement over the chosen annealing window, noting that the infidelity tends to zero over longer annealing times~\cite{SMref}. In regime (II) the transition probability is dramatically reduced and performs oscillations toward the end of the protocol, which are damped out for larger times; see SM~\cite{SMref}. Correspondingly, the purity reaches values close to unity. During the protocol, the purity is reduced at the anticrossing $(t=0)$, showing that entanglement between the qubit and spectator is instrumental in achieving high fidelity at the end of the transfer. In this respect, the protocol exhibits characteristic features of the adiabatic passage \cite{Benseny_Molmer2021_AdiabaticTheoremRevisited}. 

Figure \ref{fig:4} visualizes the central result of this work. It displays the infidelity and purity as functions of $x_0/g$ and $\omega_c/x_0$ over a relatively small annealing window. The figure is obtained for sweep rates $\epsilon$ at which the LZ transfer's asymptotic infidelity exceeds 0.45. As expected from the analysis of the spectrum in the adiabatic basis, the transfer is inefficient in the regions corresponding to regimes (I) and (III). In contrast, in regime (II) the infidelity falls below 0.05, with dark blue regions signifying values below $10^{-3}$. The latter occur along a stripe at $\omega_c/x_0\lesssim 1$ and in islands distributed along an optimal value of $\Delta_0$. Their size shows that the efficiency is robust against fluctuations of 10\% of the parameter values \cite{SMref}.
The behavior at the stripe is somehow trivial:
Here, the effect of the spectator is simply to renormalize the minimal gap of the bare LZ model to $g'=g+ x_0\approx x_0$. Remarkably, the protocol works best at the optimal value of $\Delta_0$ when $\omega_c$ is of the order of $x_0$.  In this regime, the qubit and spectator are in the ultra-strong coupling regime and become entangled during the evolution, indicating that the quantum nature of the spectator is essential to achieve high transfer efficiencies. Based on these insights, we denote by \textit{interference-assisted superadiabacity} (IAS) the high-fidelity transfer induced by the quantum field, emphasizing the role of quantum coherence and entanglement in this dynamics. 

 We have varied the qubit coupling in Eq.~\eqref{Eq:Hint} and verified that IAS occurs independently of the choice of the Pauli matrix $\hat\sigma_x$ or $\hat\sigma_y$. This indicates that IAS emerges from different mechanisms than the counteradiabatic protocol of Ref.~\cite{Berry_2009}, which instead strictly requires a classical control field along $\hat \sigma_y$ for perfect cancellation. In our case, IAS is due to interference between the different paths corresponding to the branches in the adiabatic basis, as visible in the evolution of the purity, Fig.\ \ref{fig:3}.  Further analysis indicates that IAS is relatively robust against the spectrum of the ancillary system: In the SM~\cite{SMHO} we show that the results of Fig.\ \ref{fig:4} are qualitatively reproduced when the spectator is a harmonic oscillator. Interestingly, weak dissipation on the spectator can reduce the sensitivity to parameter choices and speed up convergence to the asymptotic value~\cite{Menu_2022, SMref}.

\noindent\textit{Discussion.}---IAS could be realized in solid-state \cite{Forn-Diaz:2019,Stehli_2023,Pita_Vidal_2023} or CQED platforms \cite{Periwal_etal2021_ProgrammableInteractionsEmergent}, which offer the possibility to selectively couple a qubit with a spectator field and allow for coupling strengths in the ultra-strong coupling regime. The infidelity could be further reduced by optimizing the schedule for the qubit sweep, e.g.\ by implementing local adiabatic evolution as in Ref.\,\cite{Roland_2002}. The IAS protocol could be key for efficient annealing using cavities or central spins \cite{Anikeeva_2021,Pino_2020}: tuning the spectator's frequency controls when the annealing gap is crossed, lifting the demanding requirement of knowing the location of the gap in the parameter landscape. Moreover, the robustness against fluctuations of the parameters of the spectator field relaxes the requirements on the knowledge about the gap's size. Preliminary calculations with strings of few qubits show signatures of high-fidelity quantum state transfer. The IAS strategy could be extended to quantum gates \cite{Rao_Molmer2014_RobustRydberginteractionGates,Jandura:2024} and is in line with the insights of recent works \cite{Cai_2024,Puente_2024,dauria2025decoherencecancellationnoiseinterference, Setiawan_2021, Setiawan_2023}, pointing toward the advantage of autonomous quantum protocols for quantum supremacy. 

\bigskip

\noindent\textit{Acknowledgments.}---The authors thank L. Giannelli for insightful remarks and are grateful to R. Fazio, E. Shimshoni, C.P. Koch, Y. Gefen, I. Gornyi and J. Schachenmayer for stimulating comments. We also thank G. Anfuso for checking the numerical results. This work was partially funded by the Deutsche Forschungsgemeinschaft (DFG, German Research Foundation) – Project-ID 429529648 – TRR306 QuCoLiMa (``Quantum Cooperativity of Light and Matter"), by the German Ministry of Education and Research (BMBF) via the Project NiQ (Noise in Quantum Algorithms), and within the QuantERA II Programme (project ``QNet: Quantum transport, metastability, and neuromorphic applications in Quantum Networks"), which has received funding from the EU's Horizon 2020 research and innovation programme under Grant Agreement No. 101017733, as well as from the Deutsche Forschungsgemeinschaft DFG (Project ID 532771420). In Strasbourg, R.M. acknowledges support by the ERC Consolidator project MATHLOCCA (Grant nr.~101170485), and by the French National Research Agency under the France 2030 program ANR-23-PETQ-0002 (PEPR project QUTISYM) as well as the Investments of the Future Program project ANR-21-ESRE-0032 (aQCess).

\emph{Data availability.\textemdash}The data and code that support the findings of this article are openly available at \cite{data_reference}.

\bibliography{references}

\clearpage 
\onecolumngrid
\allowdisplaybreaks
\setcounter{equation}{0}
\setcounter{figure}{0}
\setcounter{table}{0}
\setcounter{section}{0}
\renewcommand{\theequation}{S\arabic{equation}}
\renewcommand{\thefigure}{S\arabic{figure}}
\renewcommand{\thetable}{S\arabic{table}}
\renewcommand{\thesection}{\Roman{section}}
\setcounter{secnumdepth}{3}               
\makeatletter
\renewcommand\@seccntformat[1]{\csname the#1\endcsname.\quad}
\makeatother

\setcounter{page}{1}
\renewcommand{\thepage}{S\arabic{page}}

\begin{center}
  \textbf{\Large Supplemental material for\\``Shortcuts to adiabaticity with a quantum control field''}\\[12pt]
  \normalsize Emma C. King \orcidlink{0000-0002-6696-3235},$^{1}$ Giovanna Morigi \orcidlink{0000-0002-1946-3684},$^{1,2}$ and Rapha\"el Menu \orcidlink{0000-0001-7641-9922}$^{1,3}$\\[4pt]
    {\itshape \small
  $^{1}$Theoretische  Physik,  Universit\"at  des  Saarlandes,  D-66123  Saarbr\"ucken,  Germany\\
  $^{2}$Center for Quantum Technologies (QuTe), Saarland University, Campus, 66123 Saarbr\"ucken, Germany\\
  $^{3}$CESQ/ISIS (UMR 7006), CNRS and Universit\'{e} de Strasbourg, 67000 Strasbourg, France}
  
  {\small (Dated: February 23, 2026)}
\end{center}

\onecolumngrid
\tableofcontents

\section{Spectral properties of the full Hamiltonian}\label{sec:SM:energy_spectrum}

The interference and entanglement generated by the coupling to the quantum field are key mechanisms behind our protocol: they reduce the annealing infidelity while keeping the total annealing time short. In practice, this provides an autonomous shortcut to adiabaticity for our system, achieved without introducing engineered counterdiabatic driving terms in the Hamiltonian. As seen in the main text, the fidelity of the coupled system exceeds that obtained by annealing the isolated two-level Landau-Zener (LZ) system. Interestingly, the effects of the system-field coupling on the dynamics are reflected in the structure of the full Hamiltonian spectrum. A spectral analysis reveals three qualitatively distinct regimes in which interference and entanglement act differently; see main text. These regimes determine whether the coupling enhances adiabatic following, degrades it, or instead corrects diabatic excitations accumulated during the sweep. In what follows we derive the relevant spectral features of the full Hamiltonian and explicitly relate them to the dynamical behavior observed in regimes (I)-(III) described in the main text. 

The spectator leads to a doubling of the two branches of the LZ model. Depending on the parameter values, the spectator further modifies the spectrum by opening the central gap at $t=0$ and creating additional anticrossings or ``side gaps'' at $t\neq0$. To quantify this, we exactly diagonalize the full Hamiltonian, giving four eigenvalues:
\begin{equation}\label{eq:lambda}
	\lambda=\pm\tfrac{1}{2}\sqrt{S\pm 2R}\,,
\end{equation}
where
\begin{equation}
	S =(\epsilon t)^2+\omega_c^2+g^2+4x_0^2\,,\qquad R	=\sqrt{(\epsilon t)^2\,\omega_c^2+g^2\,\omega_c^2+4g^2 x_0^2}\,. \label{eq:SandR}
\end{equation}
For distinguishability, we label the four eigenvalues by $\lambda_i$, $i=1,2,3,4$, such that $\lambda_{1}=+\tfrac{1}{2}\sqrt{S+2R},\; \lambda_{2}=+\tfrac{1}{2}\sqrt{S-2R},\;\lambda_{3}=-\tfrac{1}{2}\sqrt{S+2R},\;\text{and}\; \lambda_{4}=-\tfrac{1}{2}\sqrt{S-2R}$. The gaps in the energy spectrum are then given by
\begin{align}
			\Delta_{12}&=\lambda_1-\lambda_2=\tfrac{1}{2}(\sqrt{S+2R}-\sqrt{S-2R} ),\\
			\Delta_{23}&=\lambda_2-\lambda_3=\tfrac{1}{2}(\sqrt{S-2R}+\sqrt{S+2R}),\\
			\Delta_{24}&=\lambda_2-\lambda_4=\sqrt{S-2R},\\
			\Delta_{13}&=\lambda_1-\lambda_3=\sqrt{S+2R}.
\end{align}
Note that other energy differences are either negatives of the above or trivial combinations. We now consider the middle gap
		\begin{equation}
			\Delta_{24}(t)=\sqrt{S-2R}\,\ge 0
		\end{equation}
and locate the extrema by solving $\tfrac{\mathrm d}{\mathrm d t}\Delta_{24}(t)=0$ for time $t$. After some algebra, we arrive at three solutions
\begin{equation}\label{eq:gap_times}
    t_{\mathrm{ALC}}=0\,, \qquad t_{\mathrm{sg}}=\pm\frac{1}{\varepsilon}\sqrt{\,\omega_c^2-g^2- 4g^2\,x_0^2/\omega_c^2\,}\,.
\end{equation}
The first solution, $t=t_\mathrm{ALC}$, corresponds to the time at which the isolated Landau-Zener system passes through the avoided level-crossing. In the full description, the gap at $t=t_\mathrm{ALC}$ is given by
\begin{equation}\label{eq:exact_gap_t0}
    \Delta(t_\mathrm{ALC})= \sqrt{g^2+\omega_c^2 + 4x_0^2 - 2\sqrt{g^2(\omega_c^2+4 x_0^2)}} = \left\vert g-\sqrt{\omega_c^2+4x_0^2}\right\vert\equiv \vert g-\Delta_0\vert\,.
\end{equation}
Note that $x_0$ leads to an effective renormalization of the gap, allowing for more rapid passage through the level-crossing. 

\begin{figure}[t]
    \centering
    \includegraphics[width=0.495\linewidth]{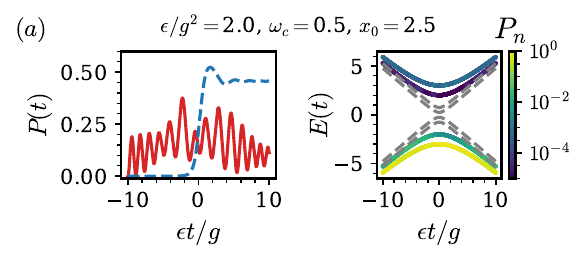}
    \includegraphics[width=0.495\linewidth]{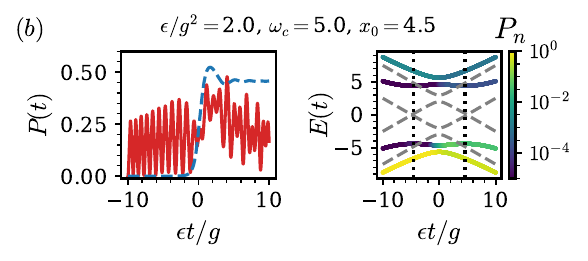}
    \includegraphics[width=0.495\linewidth]{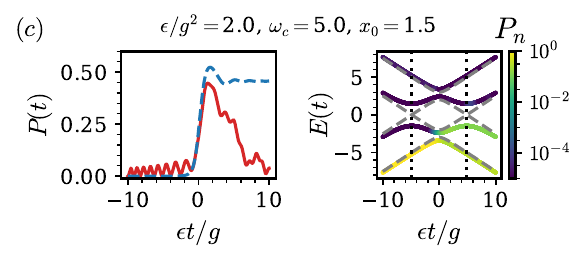}
    \includegraphics[width=0.495\linewidth]{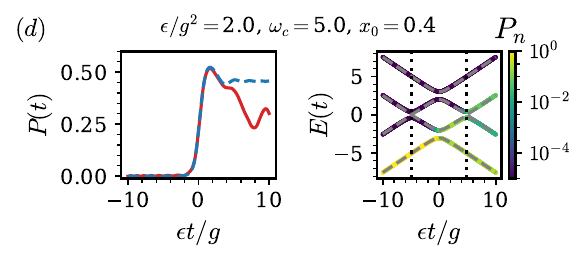}
    \caption{(a)-(d) (Left): Dynamics of the transition probability $P(t)$ of the Landau-Zener system after tracing out the quantum field. Solid red curves denote results for finite system-spectator couplings, $x_0\neq0$. The dashed blue curve illustrates the case where the system and spectator qubit are decoupled, $x_0=0$, and is only shown for reference. Parameter values are specified above the respective subfigures. We consistently fix $g=1$. (a)-(d) (Right): Energy branches $E_n(t)$, $n=1,\,2,\,3,\,4$, of the full system. The branches are colored according to the instantaneous population of each eigenstate, where the population of the eigenstate $\vert n(t)\rangle $ is computed as $P_n(t)=\vert \langle n(t)\vert\Psi\rangle_t\vert^2$. Dashed gray bands denote the decoupled case $(x_0=0)$. Dotted black vertical lines indicate the times at which the side gaps occur, as predicted by Eq.~\eqref{eq:gap_times}. (a) $\omega_c<g$, since $g=1$. Side gaps vanish, and we observe an effective renormalization of the central gap. (b) $\omega_c>g$, $t_\mathrm{sg}<t_f$, and $x_0/g\gg1$. The conditions on the spectator frequency and final anneal time for the side gaps to appear are satisfied; however, the strong coupling results in oscillations leading to transitions out of the system's instantaneous ground state and the population of the highest energy branch. (c) The coupling $x_0$ is now of the order of $g$. Additional anticrossings lead to energy branches corresponding to the instantaneous ground state $\vert-\rangle_t$ being populated, reducing the infidelity. (d) The coupling $x_0$ is small relative to $g$. Diabatic transitions occur at the second side gap, resulting in population transfer out of the branches coinciding with the instantaneous ground state $\vert-\rangle_t$ of the system.     }
    \label{fig:e_spectrum}
\end{figure}

The second solution $t_\mathrm{sg}$ \eqref{eq:gap_times} provides the times at which the ``side gaps'' appear in the energy spectrum; refer to Fig.~2 of the manuscript and Fig.~\ref{fig:e_spectrum}. The frequency $\omega_c$ controls the location of these two new anticrossings, shifting the crossing point either away from or toward the central crossing point. If the frequency and coupling strength are chosen such that $t_\mathrm{sg}>t_f$, the side gaps lie outside the annealing time window, and we enter the so-called \textit{regime (III)} (see Fig.~4 of the main text, with the dashed boundary between regime (II) and (III) demarcating the points at which $t_\mathrm{sg}=t_f$). Here performance notably deteriorates. Now consider the case when $t_\mathrm{sg}<t_f$. Provided $|\omega_c|\gtrsim \sqrt{g\Delta_0}$ with $\Delta_0=\sqrt{4x_0^2+\omega_c^2}$, the gaps at time $t=t_\mathrm{sg}$ have a magnitude
\begin{equation}
    \Delta(t_\mathrm{sg}) = x_0\sqrt{1-g^2/\omega_c^2}\,,
\end{equation}
which scales with $x_0$. There is a delicate interplay between $x_0$ and the gap $g$. Setting $x_0/g\gg1$ yields large side gaps relative to the gap $g$ of the isolated LZ model. While this allows adiabatic passage through these additional anticrossings, it leads to oscillatory behavior in the eigenstates and populates the uppermost energy branch; see Fig.~\ref{fig:e_spectrum}(b). Ideally, we require the gap of the first anticrossing to be small, with a large gap at the final anticrossing $t=+|t_\mathrm{sg}|$, such that transitions out of the subspace of the instantaneous ground state of the system $\vert -\rangle_t$ are suppressed. On the other hand, fixing $x_0/g\ll1$ leads to diabatic transitions at the anticrossing $t=+t_\mathrm{sg}$ after the central gap at $\epsilon t/g=0$. Depending on how severely the adiabatic theorem is violated here, we can still partially recover some populations in the second-lowest energy branch, corresponding to a reduction in the infidelity. Refer to Fig.~\ref{fig:e_spectrum}(d). Ideally, $x_0$ is of the order of the gap $g$; the spectator frequency $|\omega_c|<\epsilon t_f$ is then tuned to coincide with a destructive interference fringe, thereby reducing the infidelity (Fig.~\ref{fig:e_spectrum}(c)).  When $|\omega_c|\lesssim \sqrt{g\Delta_0}$ there is no real crossing at $t_\mathrm{sg}$: the global minimum lies at $t=0$ and has a magnitude given by Eq.~\eqref{eq:exact_gap_t0}, as shown in Fig.~\ref{fig:e_spectrum}(a).

\section{Long-time dynamics}\label{SM:sec:LTD}
	In the isolated Landau–Zener (LZ) model the transition probability dynamics $P(t)$ is inherently oscillatory: large-amplitude oscillations appear in the vicinity of the avoided crossing and persist after the crossing has been traversed, eventually ``damping out'' to a fixed asymptotic value at later times. In our protocol, we intentionally operate in short, finite-time windows deep in the non-adiabatic regime, i.e.\ on timescales for which the instantaneous energy gap is small compared with the coupling-induced energy scale, so that the adiabatic theorem is violated. These short time windows are necessary to realize a shortcut to adiabaticity, keeping total runtime (and exposure to decoherence) small. However, a direct consequence is that both the transition probability $P(t)$ and the purity $\gamma(t)$ remain strongly time dependent at the end of the anneal $t=t_f$, and have not relaxed to their asymptotic values by $t_f$. Hence, we expect pronounced finite-time oscillations in the observables when the shortcut is exploited. This is observed in Fig.~3 of the main text. 

    \begin{figure}[b]
    \centering
    \includegraphics[width=\textwidth]{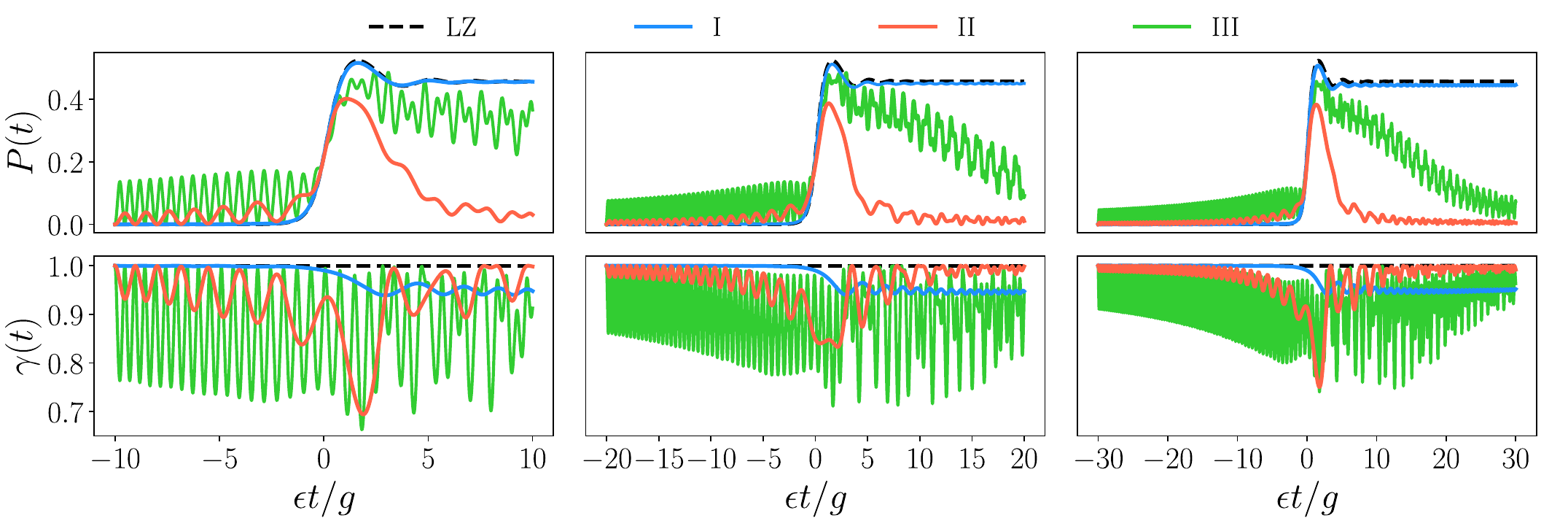}
    \caption{Evolution of the transition probability $P(t)$ and the purity $\gamma(t)$ as a function of time for the qubit-field dynamics for $\epsilon = 2g^2$. Three different choices of $x_0$ and $\omega_c$ are shown, illustrating the regimes of the annealing dynamics: (I) $x_0 = g/10$ and $\omega_c = x_0$, (II) $x_0 = 5g/4$ and $\omega_c = 5x_0/2$, (III) $x_0 = 5g$ and $\omega_c = 3x_0$. The black dashed line represents the Landau-Zener case ($x_0=0$). The qubit-field system is initialized at time $t_i$ in state $\vert\Psi\rangle_0 = \vert -\rangle_{t_i}\otimes\vert\downarrow\rangle $ and evolved up until time $t_f$. These results are displayed for three choices of initial/final times $t_{f/i} =\pm 10 g/\epsilon$, $\pm 20 g/\epsilon$ or $\pm 30 g/\epsilon$.}
    \label{fig:sm1}
\end{figure}

    Extending the total annealing time reduces the oscillation amplitudes in regimes (I) and (II) (compare shorter \textit{vs} longer annealing durations $2t_f$ in Fig.~\ref{fig:sm1}). Further reductions in the infidelity characterize longer-time dynamics in regime (III), which can be viewed as an effective expansion of parameter regime (II); see dark blue region in Figs.~\ref{fig:sm2} and \ref{fig:sm3}. Consequently, extending the anneal time could be used as an alternative approach to our simple optimization strategy, the latter of which requires stopping at the precise time instance at which the infidelity is minimized (see Sec.~\ref{sec:SM:optimization} for details). Thus, there is a clear trade-off: operating at short times (small $t_f$) is required to take full advantage of the shortcut to adiabaticity, but comes at the price of strong residual oscillations in $P(t)$ and $\gamma(t)$. In contrast,  increasing $t_f$ damps the oscillations, but forfeits the fast shortcut. In Sec.~\ref{sec:SM:interference_initial_state} we elaborate on another approach to combat these undesirable oscillations in the dynamics, tracing their origin, in part, to the choice of initial state and the consequent emergence of interference and entanglement effects.

	We close with a final remark on the long-time dynamics where finite-time effects are suppressed. Notice that the beneficial reduction of the infidelity produced by coupling to the quantum field is preserved at late times in regime (II), and can even be enhanced in regime (III) (Fig.~\ref{fig:sm1}). This implies that the field-induced suppression of nonadiabatic transitions is not solely a transient destructive interference effect that persists at short times, but also reduces the long-time transition probability. For a comparison, see, for example, Figs.~\ref{fig:sm2} and \ref{fig:sm3} where the annealing time window is $t_i=-t_f=10g/\epsilon$ and $t_i=-t_f=30g/\epsilon$, respectively.

\begin{figure}[h]
    \centering
    \includegraphics[width=0.99\linewidth]{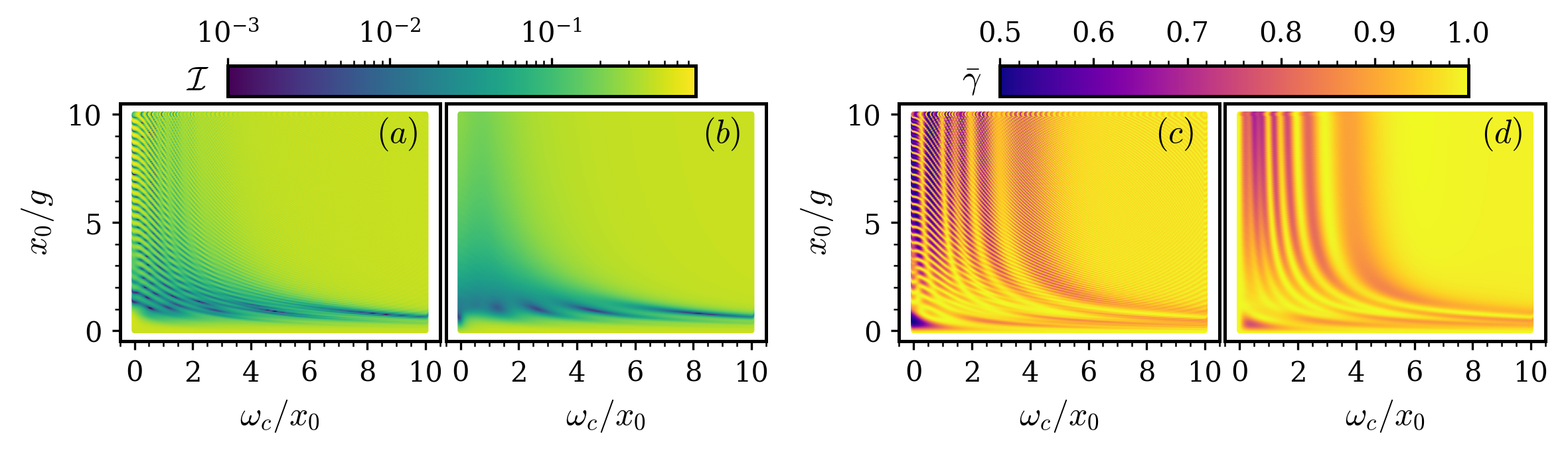}
    \caption{(a) and (c): Infidelity $\mathcal{I}=P(t_f)$ and purity $\bar{\gamma}=\gamma(t_f)$, respectively, at the final time $t_f$ of the anneal when the system is initialized in the product state $\vert\Psi\rangle_0=\vert -\rangle_{t_i}\otimes|\downarrow\rangle$ at time $t_i$. (b) and (d): As in (a) and (c), but the initial state $\vert \Psi\rangle_0$ is chosen to be the lowest-energy eigenstate of the full Hamiltonian $\hat{H}$ at time $t_i$. The effect of choosing an entangled state that is an eigenstate of $\hat{H}$ \textit{vs} a product state is a ``smoothing out'' of the interference fringes.}
    \label{fig:density_plot_product_vs_fullGS}
\end{figure}

\begin{figure}[h]
    \centering
    \includegraphics[width=0.9\linewidth]{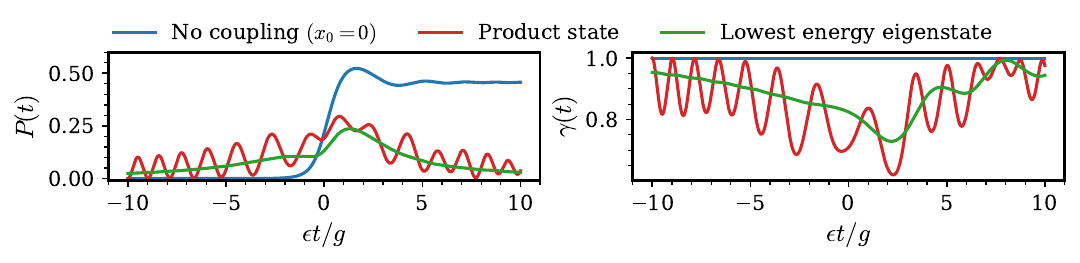}
    \includegraphics[width=0.9\linewidth]{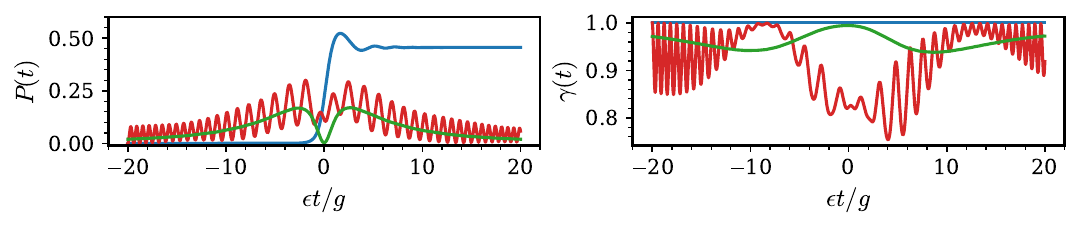}
    \caption{Evolution of the transition probability $P(t)$ (left) and the purity $\gamma(t)$ (right) as a function of time for the qubit-field dynamics for two different choices of initial states: product state $\vert\Psi\rangle_0 = \vert-\rangle_{t_i}\otimes\vert\downarrow\rangle$ (red) and instantaneous ground state of the full Hamiltonian (green). For reference, the dynamics without the spectator ($x_0=0$) is shown by the blue curve. Parameters are $x_0=\omega_c=2g$, $\epsilon=2g^2$ and $t_f=-t_i=10g/\epsilon$ in the top panel and $x_0=3g$, $\omega_c=x_0/6$, $\epsilon=2g^2$ and $t_f=-t_i=20g/\epsilon$ in the bottom panel.}
    \label{fig:cuts_product_vs_fullGS}
\end{figure}

\section{Influence of the initial state}\label{sec:SM:interference_initial_state}

\begin{figure}[t]
    \centering
    \includegraphics[width=0.45\linewidth]{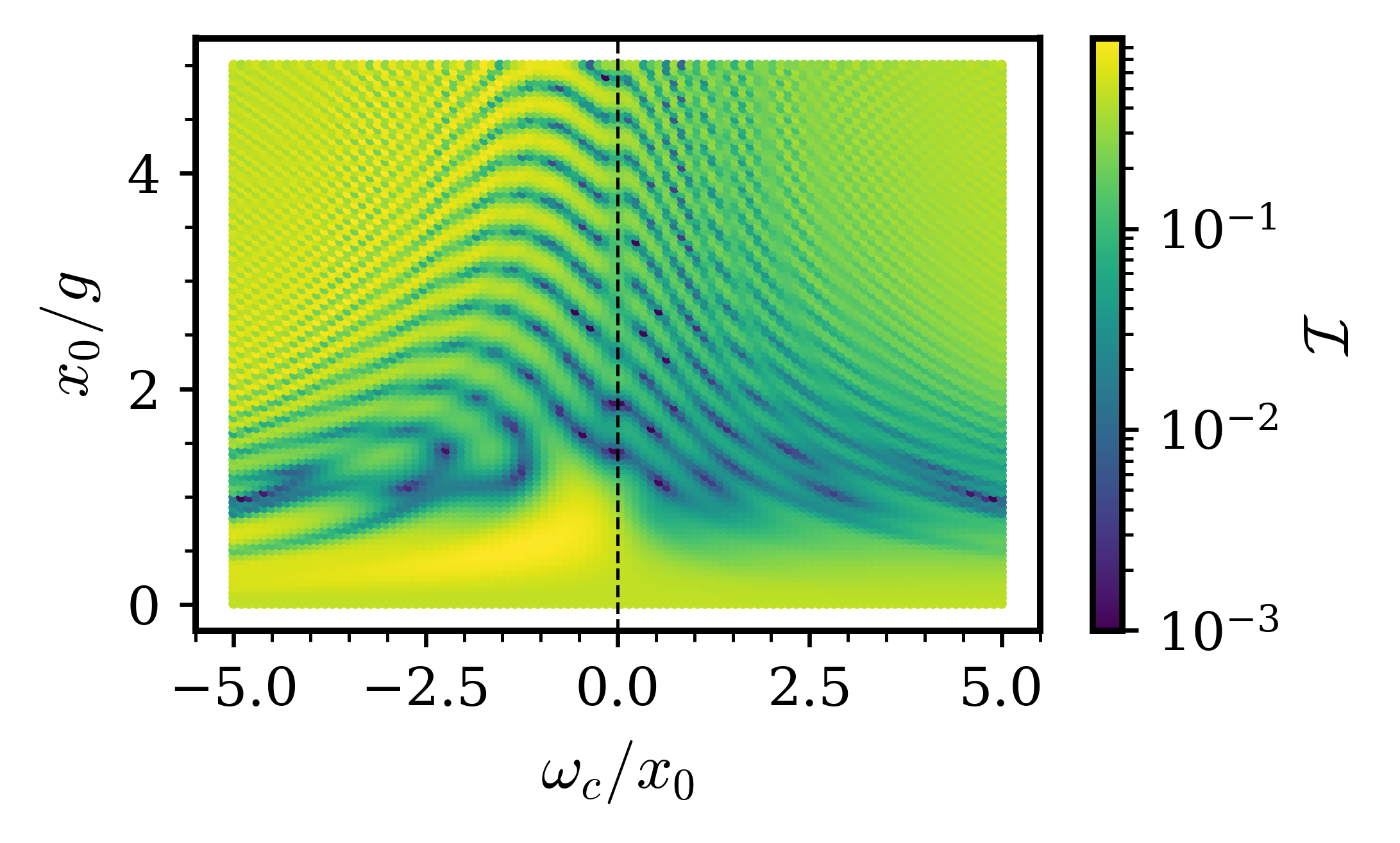}
    \includegraphics[width=0.45\linewidth]{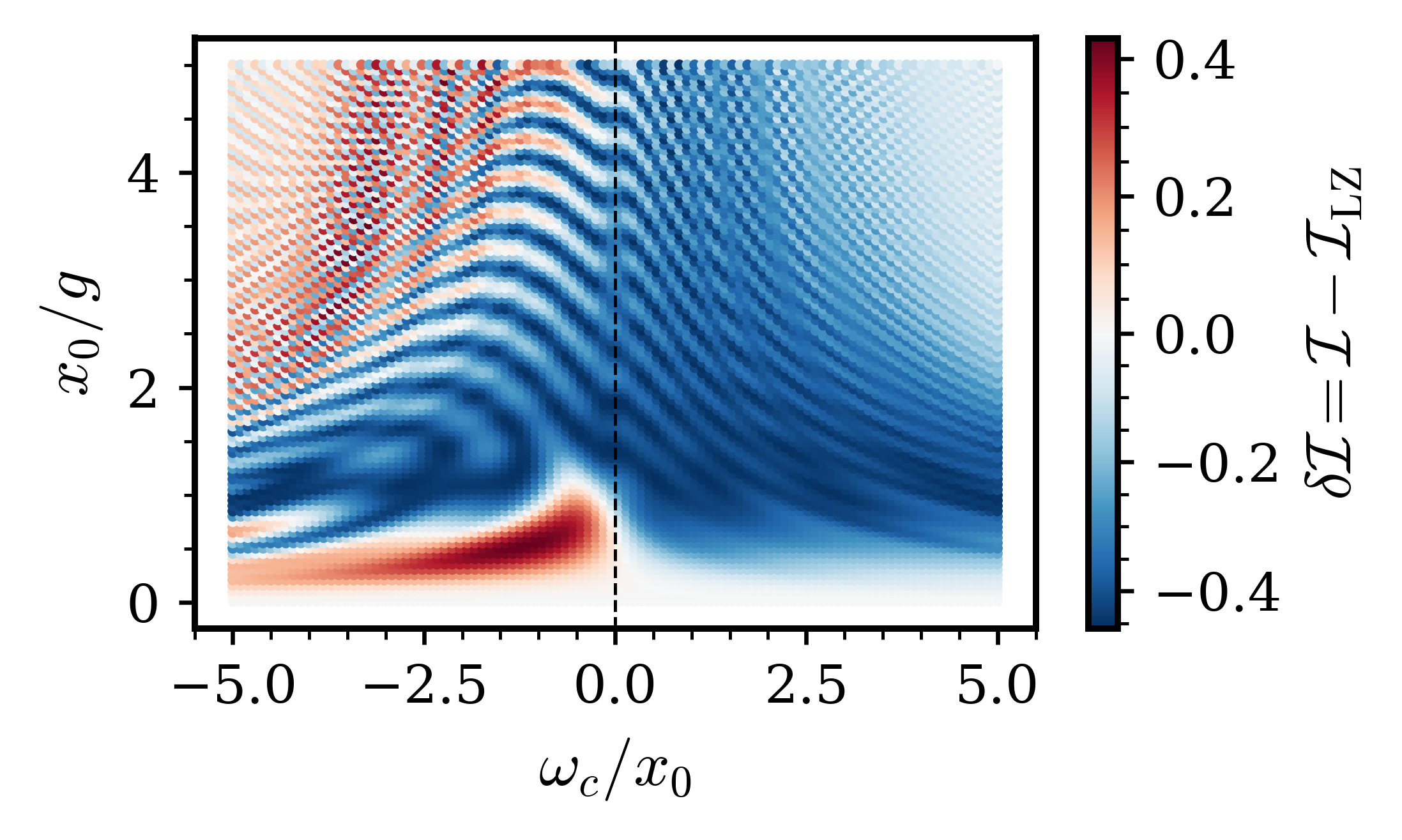}
    \caption{(Left) The transition probability $\mathcal{I}=P(t_f)$ at the final time $t_f=10g/\epsilon$ of the anneal as a function of the rescaled coupling strength $x_0/g$ and spectator frequency $\omega_c/x_0$. We set $\epsilon=2g^2$ and fix the initial state at time $t_i=-t_f$ to be the product state $\vert\Psi\rangle_0=\vert-\rangle_{t_i}\otimes \vert\downarrow\rangle$, with $\vert-\rangle_{t_i}$ the instantaneous ground state of the system Hamiltonian and $\vert\downarrow\rangle$ the ground state of the spectator (field) Hamiltonian $\hat H_f$ with $\omega_c>0$. Tuning the frequency $\omega_c$ to take negative values (while keeping the initial state $\vert\Psi\rangle_0=\vert-\rangle_{t_i}\otimes \vert\downarrow\rangle$ fixed) therefore corresponds to the spectator initially being excited. Larger regions of yellow shading to the left of the dashed vertical line (at $\omega_c/x_0=0$) indicate larger infidelities with respect to the case of initializing the spectator in its ground state. (Right) Infidelity results of the left panel, but now deducting the infidelity $\mathcal I_{\mathrm LZ}$ obtained with the isolated Landau-Zener model. Blue regions ($\delta\mathcal I<0$) indicate that our protocol outperforms the isolated LZ case, while red regions ($\delta\mathcal I>0$) signify a deterioration in the achievable fidelity due to the coupling to the quantum field. Note that, independent of the coupling and frequency choices, for a field that is initialized in its ground state (here $\omega_c>0$), we always obtain reduced infidelities compared to the isolated LZ system.  }
    \label{fig:pos_vs_neg_frequencies}
\end{figure}

Results of the main text, see Fig.~4, suggest an underlying interference effect in the dynamics, resulting in clear fringes of higher and lower infidelity and purity. In addition to this, we obtain further interference, attributed to the `oscillations' between the eigenstates of the full system (system+spectator).  The latter is due to the choice of initial state being a product state, with each subsystem, the LZ two-level system and the spectator qubit, being initialized in their respective instantaneous ground states, i.e.\ $\vert\Psi\rangle_0 = \vert -\rangle_{t_i}\otimes\vert\downarrow\rangle$. One can reduce this effect by initializing the state of the full system in a state that is close to an instantaneous eigenstate of the Hamiltonian $\hat{H}$. This `smoothes out' the small interference bands in both the infidelity and purity; refer to Fig.~\ref{fig:density_plot_product_vs_fullGS}. An additional advantage is that some of the oscillations observed in the dynamics (Fig.~\ref{fig:sm1}) may be suppressed, but generally come at the expense of a less pure final state compared to the optimized dynamics starting from a product state; see Fig.~\ref{fig:cuts_product_vs_fullGS}. As a final remark, we stress that when initializing the system in a product state, choosing the spectator qubit to be spin-up or spin-down does make a difference. This can be benchmarked numerically by conveniently changing the spectator frequency and keeping the initial state $\vert\downarrow\rangle$ fixed. For $\omega_c<0$ (spectator is initially excited), there is a reduction in performance, with performance being quantified by the lowest attainable infidelity. Despite this, we find that for appropriately tuned spectator frequencies and coupling strengths, we can still achieve low infidelities. See  Fig.~\ref{fig:pos_vs_neg_frequencies} and the discussion in the caption.

\begin{figure}[t]
    \centering
    \includegraphics[width=0.85\linewidth]{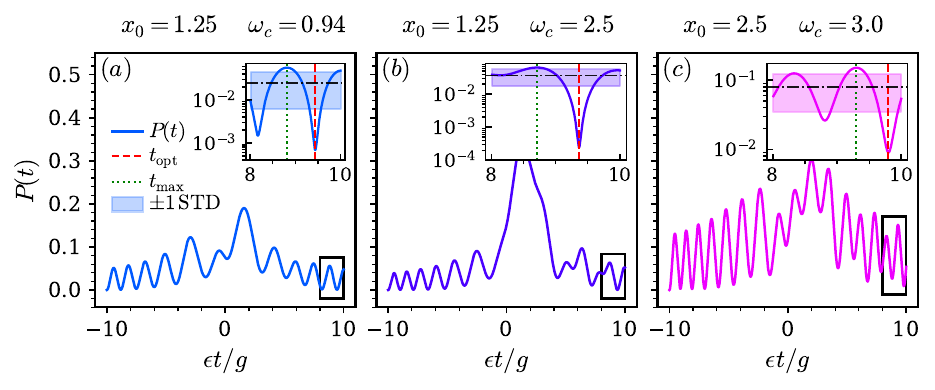}
    \caption{Dynamics of the transition probability $P(t)$ for different parameter values, as specified above subfigures (a)-(c) . Black boxes enclose the time interval $\epsilon t/g\in[0.8\,\epsilon t_f/g,\epsilon t_f/g]$ in which we select the smallest attainable infidelity in our optimized protocol, which we denote $\mathcal I_\mathrm{opt}$. Insets enlarge this final time window (10\% of the total annealing time), showing the minimum infidelity $\mathcal I_\mathrm{opt}$ at  $t_{\rm opt}$ (red dashed line) and maximum infidelity at $t_{\rm max}$ (green dotted line). The mean $\bar{\mathcal I}$ is shown by the black dot-dashed curve, with one standard deviation indicated by the shaded color.  }
    \label{fig:sm5}
\end{figure}
\begin{figure*}[t]
    \centering
    \includegraphics[width=0.9\textwidth]{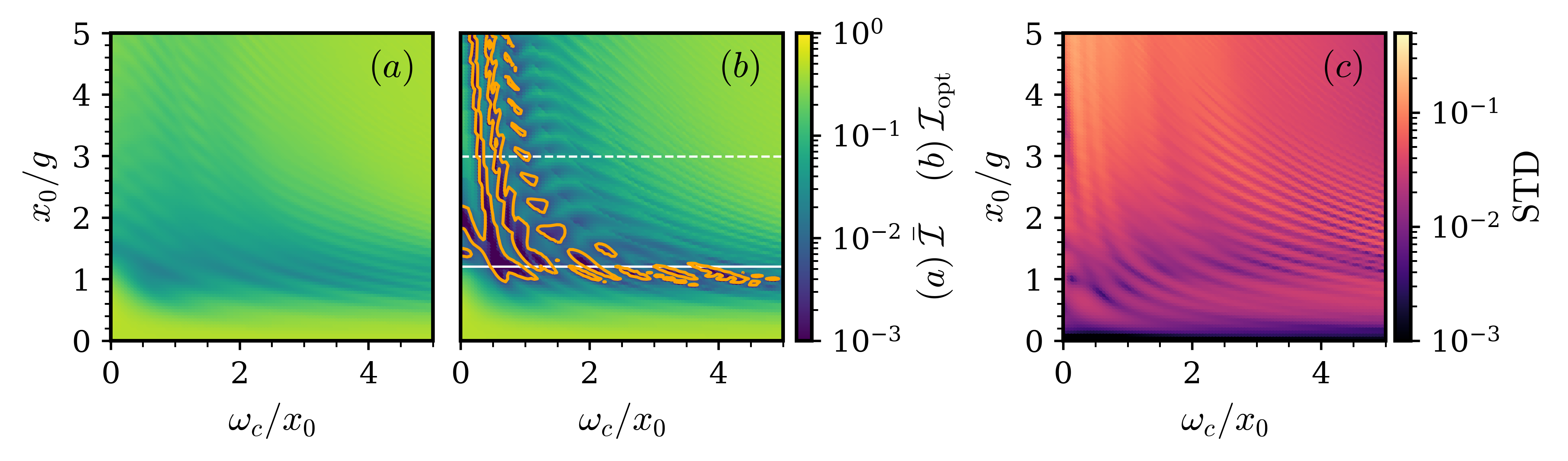}
    \caption{(a) Mean infidelity $\bar{\mathcal{I}}$, averaged over the final 10\% of the anneal time, shown as a function of the parameters $x_0/g$ and $\omega_c/x_0$. (b) Optimized infidelity $\mathcal{I}_\mathrm{opt}$, defined as the minimum infidelity attained within the final 10\% of the anneal time (see Fig.~4 in the main text). Orange contours demarcate regions where the infidelity drops below $5\times 10^{-3}$. Slices along the white horizontal lines are shown in Fig.~\ref{fig:cuts}. (c) Standard deviation (STD) of the infidelity over the same final-time window, computed as $\sqrt{\sum_j |\mathcal I_j - \bar{\mathcal{I}}|^2/(N-1)}$, where $\mathcal{I}_j$ denotes the infidelity at data point $j$ in the window, $\bar{\mathcal{I}}$ is the mean (see (a)), and $N$ is the number of data points. All figures use fixed parameters $\epsilon = 2g^2$ and an annealing time window $t_f = -t_i = 10g/\epsilon$. }
    \label{fig:sm2}
\end{figure*}

\section{Optimization of protocol and robustness}\label{sec:SM:optimization}

Due to the pronounced oscillatory behavior of the transition probability (see Sec.~\ref{SM:sec:LTD} and discussion therein), we ``optimize'' the termination time of the annealing sweep to obtain the lowest achievable infidelity. Concretely, for each set of system parameters we search the interval $t\in[0.8\,t_f,t_f]$ and select the (closest) local minimum of the diabatic transition probability, denoted $P(t_{\rm opt})$; see Fig.~\ref{fig:sm5}. To assess the degree of sensitivity to the precise end time, we compare three quantities at the end of the protocol: the unoptimized final infidelity $\mathcal{I}=P(t_f)$, the optimized infidelity $\mathcal{I}_{\rm opt}=P(t_{\rm opt})$, and the time-averaged infidelity $\bar{\mathcal{I}} \equiv \frac{1}{0.2\,t_f}\int_{0.8\,t_f}^{t_f} P(t)\,\mathrm{d}t$. The latter is simply the mean value of the infidelity over the final $10\%$ of the anneal, with the corresponding standard deviation computed over the same time interval as $\sqrt{\sum_{j} |P(t_j) - \bar{\mathcal{I}}|^2/(N-1)}$. Figure~\ref{fig:sm2} summarizes the numeric optimized and mean results.

\begin{figure*}[t]
    \centering
    \includegraphics[width=0.9\textwidth]{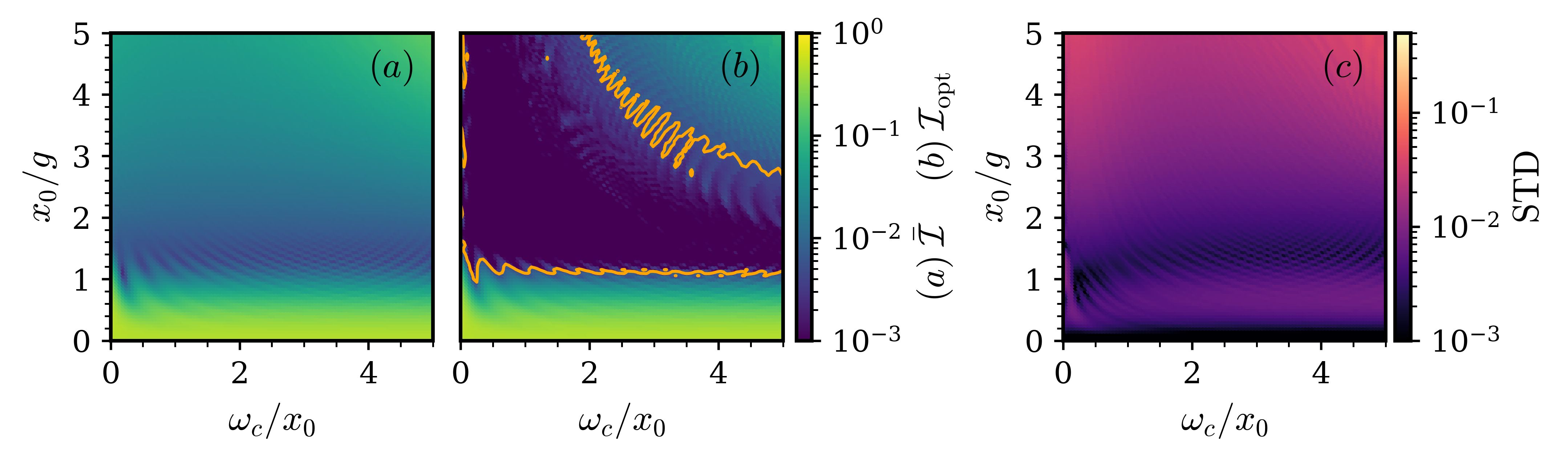}
    \caption{As in Fig.~\ref{fig:sm2}, but for a longer annealing time window $t_f = -t_i = 30g/\epsilon$.}
    \label{fig:sm3}
\end{figure*}

Two points are immediately apparent. First, the interference bands that are visible in the optimized-time density plot, Fig.~\ref{fig:sm2}(b), are smoothed out in the mean $\bar{\mathcal{I}}$, with the amplitude of the fringes/bands being worsened by roughly one order of magnitude relative to the instantaneous optimized probabilities. Despite this, the overall interference pattern remains qualitatively intact and high fidelities are still attainable in certain parameter regimes. The latter remark brings us to the second point: coupling the Landau–Zener (LZ) qubit to the spectator field both reduces the infidelity and produces an effective shortcut to adiabaticity, even if the anneal cannot be terminated exactly at the specified optimal time. With reference to Fig.~\ref{fig:sm2}, the asymptotic diabatic probability of the bare LZ problem is $\mathcal{I}_\mathrm{LZ}\approx 0.46$ for the chosen parameters. Note that the optimized coupled dynamics yield improvements up to 2 to 3 orders of magnitude in $\mathcal{I}_{\rm opt}$ (Fig.~\ref{fig:sm2}(b)). Although the mean value performs worse, we still observe improvements of up to 2 orders of magnitude in the infidelity (Fig.~\ref{fig:sm2}(a)). The reduced mean value $\bar{\mathcal{I}}$ compared to the asymptotic result of the isolated LZ model $\mathcal{I}_\mathrm{LZ}$ demonstrates that the beneficial effect of the spectator coupling in our protocol is not simply a single-time phenomenon. Performing multiple experimental runs in which the protocol end-time is allowed to vary within a finite window will still improve results, albeit inferior to results achieved when fixing the final time to $t=t_\mathrm{opt}$. As expected, longer annealing time windows provide qualitatively similar results, but with much lower infidelities overall and the expansion of regime~(II); see Fig.~\ref{fig:sm3}.

\begin{figure*}[t]
    \centering
    \includegraphics[width=0.45\textwidth]{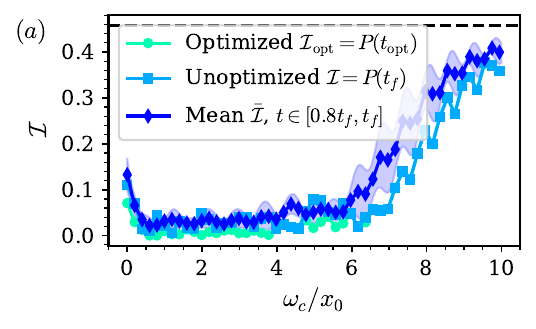}
    \includegraphics[width=0.45\textwidth]{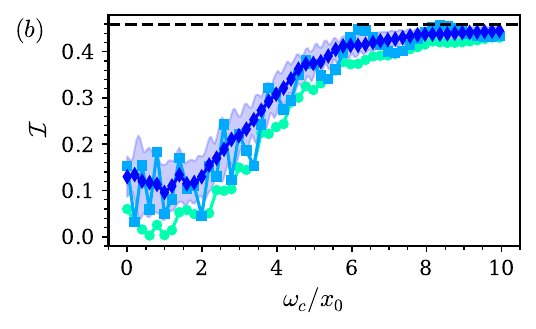}
    \caption{The optimized, unoptimized, and mean infidelity shown for two cuts $x_0/g\approx1.2$ (a) and $x_0/g\approx3$ (b) in the ($x_0/g$)-($\omega_c/x_0$) plane, as indicated by the solid line and dashed line, respectively, in Fig.~\ref{fig:sm2}(b). Standard deviation is displayed as shading around the $\bar{\mathcal I}$ curve. The dashed black line denotes the infidelity of an isolated Landau-Zener model with the same parameters. Remaining parameters are set to $\epsilon=2g^2$ and $t_f=-t_i=10g/\epsilon$.}
    \label{fig:cuts}
\end{figure*}

To provide further insight into how the different infidelities ($\mathcal{I}$, $\mathcal{I}_\mathrm{opt}$, and $\bar{\mathcal I}$) compare, in Figs.~\ref{fig:cuts}(a),(b) we plot cuts through parameter space at fixed couplings $x_0/g\approx 1.2$ and $x_0/g\approx3$, respectively; see Fig.~\ref{fig:sm2}(b). Focusing on Fig.~\ref{fig:cuts}(a), for small $\omega_c/x_0$, the interaction between the system and spectator is weak, acting as a perturbation to the isolated system. We classify this as regime (I). The mean is close to the optimized value and the standard deviation is small, indicating that the final fidelity is insensitive to small timing errors. Despite this, regime (I) is not of interest due to its characteristically low fidelities (of the order of $1-\mathcal{I}_\mathrm{LZ}$), which are attributed to the spectator field further closing the energy gap at the level-crossing (Fig.~2(a) of the main text). When $\omega_c>g$ and $t_\mathrm{sg}<t_f$, we enter regime (II); additional side gaps are formed in the energy spectrum, and the infidelity decreases (see also Sec.~\ref{sec:SM:energy_spectrum}). Here, the differences between the optimized, mean, and unoptimized results are apparent. As soon as $t_\mathrm{sg}>t_f$, the side gaps are shifted outside the time window that we consider in our finite-time protocol, entering what we call regime (III). This corresponds to a sudden noticeable increase in the infidelity. Figure~\ref{fig:cuts}(b) shows similar results, but for stronger coupling $x_0$, which gives rise to oscillatory behavior and larger discrepancies between the different curves.

\begin{figure*}[t]
    \centering
    \includegraphics[width=0.9\textwidth]{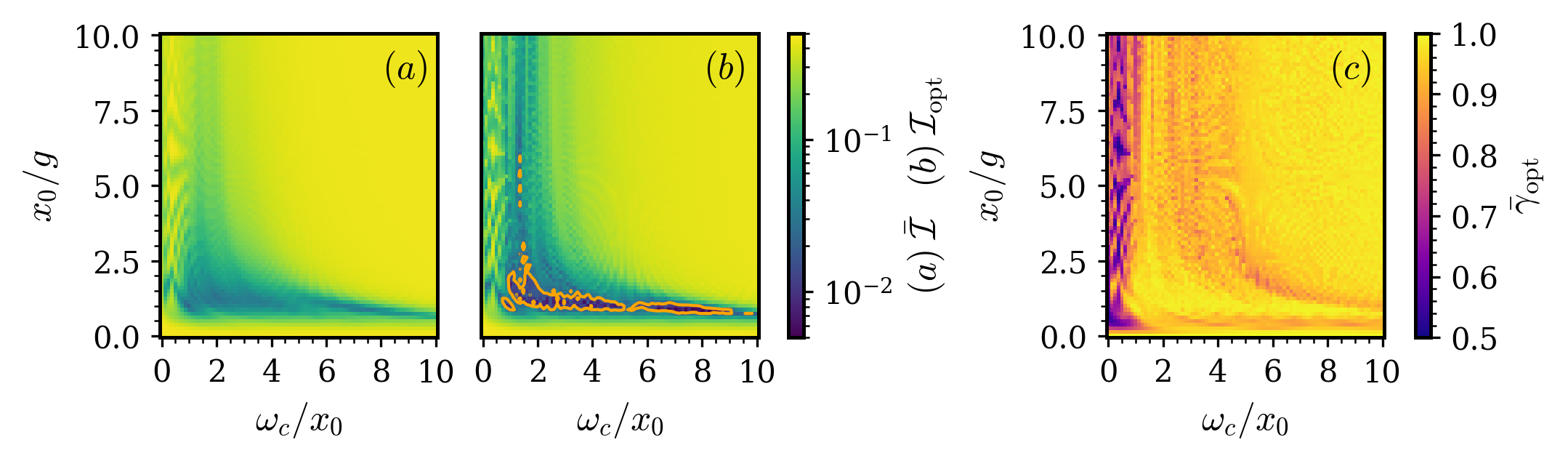}
    \caption{Adiabatic transfer in a LZ model with a harmonic oscillator as the quantum field  (Fig.~\ref{fig:sm2} shows the same results for a spectator qubit). (a) Mean infidelity $\bar{\mathcal{I}}$, averaged over the final 10\% of the anneal time, shown as a function of the parameters $x_0/g$ and $\omega_c/x_0$. (b) Optimized infidelity $\mathcal{I}_\mathrm{opt}$, defined as the minimum infidelity attained within the final 10\% of the anneal time (see Sec.~\ref{sec:SM:optimization}). The orange contours delineate regions in which the infidelity is below $2\times 10^{-2}$. (c) Purity $\gamma(t_\mathrm{opt})=\overline{\gamma}_\mathrm{opt}$ at the optimal stopping time $t_\mathrm{opt}$ where the infidelity reaches its smallest value. In (a)-(c) we fix parameters $\epsilon = 2g^2$ and choose an annealing time window $t_f = -t_i = 10g/\epsilon$. We truncate the Hilbert space of the oscillator to $n_\mathrm{max}=30$ levels.}
    \label{fig:osc1}
\end{figure*}

\begin{figure*}[t]
    \centering
    \includegraphics[width=0.9\textwidth]{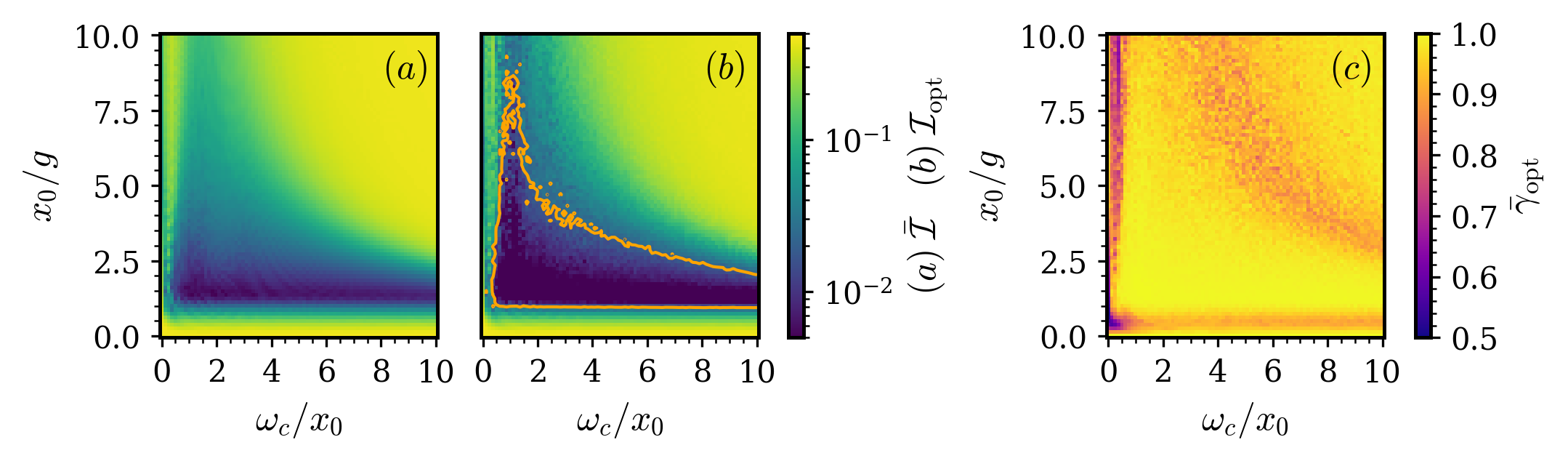}
    \caption{As in Fig.~\ref{fig:osc1}, but for a longer annealing time window $t_f = -t_i = 30g/\epsilon$.}
    \label{fig:osc_long_time}
\end{figure*}

\section{Harmonic oscillator spectator}\label{sec:HMOspectator}
We consider a natural extension of the spectator qubit considered in the main text to a quantum field with an unbounded Hilbert space, namely a harmonic oscillator. The field Hamiltonian takes the form $\hat H_f = \omega_c \hat{a}^\dagger \hat{a}$ and the interaction term in the full Hamiltonian is
\begin{equation}\label{Eq:3}
    \hat{H}_\mathrm{int} = x_0 \hat{\sigma}_x\otimes (\hat{a} + \hat{a}^\dagger)\,,
\end{equation}
where $\hat a$, $\hat a^\dagger$ are the bosonic annihilation and creation operators. As before, the initial state is a product state, comprising the instantaneous ground state of the LZ model and the vacuum state of the oscillator. In Figs.~\ref{fig:osc1}-\ref{fig:osc_long_time} we verify numerically that the phenomenology observed for the qubit spectator field applies to the case of a harmonic oscillator ancillary system. In particular, we observe the irrefutable presence of regime (II), characterized by lower infidelities and represented by the dark blue regions in Fig.~\ref{fig:osc1}. As witnessed for the qubit spectator, extending the annealing time expands regime (II); this is demonstrated in Fig.~\ref{fig:osc_long_time} for the oscillator quantum field.

\section{Dissipation: effect of a lossy spectator}\label{sec:SM:lossy_spectator}

\begin{figure}[t]
    \centering
    \includegraphics[width=0.9\linewidth]{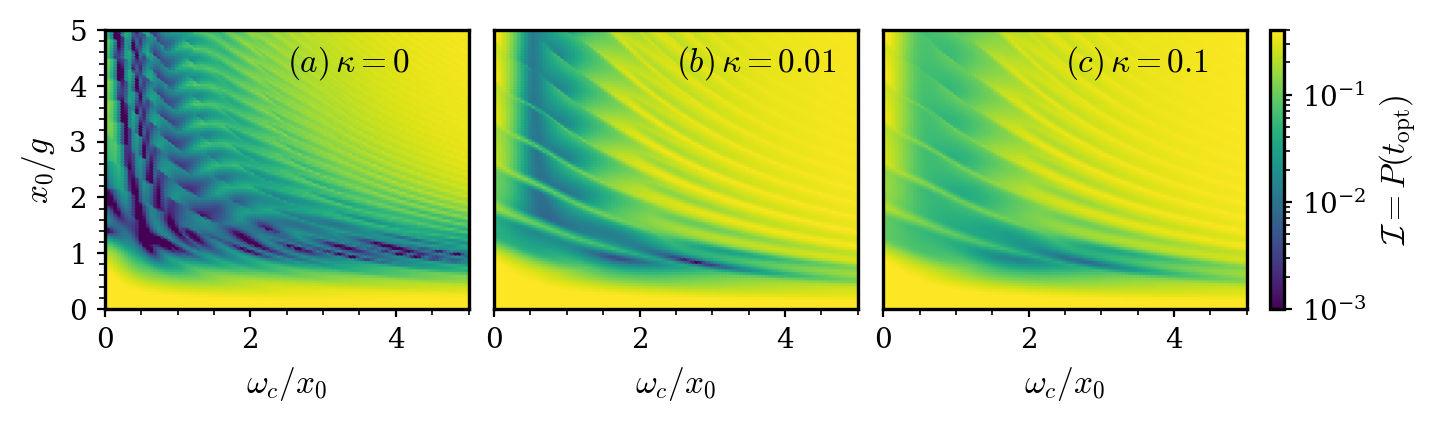}
    \caption{(a) Identical to Fig.~\ref{fig:sm2}(b), which is a subregion of Fig.~4 of the main text; shown only for reference. (b) and (c): Protocol implemented as in (a), but with weak dissipation of the spectator qubit modelled according to Eq.~\eqref{eq:master_qubit_zeroT} with $\kappa=0.01$ and $\kappa=0.1$, respectively. The transition probability $P(t)$, and therefore also the infidelity $\mathcal I=P(t_\mathrm{opt})$, is computed directly from the density matrix $\hat\chi(t)$ as $P(t) = _t\langle+\vert\mathrm{Tr}_f\{\hat\chi(t)\}\vert+\rangle_t= \mathrm{Tr}\{\hat\rho(t)\vert+\rangle_t\langle+\vert\}$ . }
    \label{fig:dissipation}
\end{figure}

In experimentally relevant implementations (for example, in cavity QED or circuit-QED architectures) incoherent processes of the spectator cannot be ignored and may appreciably modify the beneficial interference and entanglement mechanisms described in the main text. Additionally, since the dynamics of the coupled system is oscillatory, a natural question is whether weak dissipation of the spectator might in fact {\it reduce} harmful oscillations and thereby be advantageous, or whether the price paid in achievable fidelity always renders dissipation detrimental. Below we summarize the master equation we use to investigate these points and the principal conclusions. Numerical results illustrating the smoothing of the parameter dependence for small but finite damping are shown in Fig.~\ref{fig:dissipation}.

Consider the model described in the main text, but with the spectator qubit now coupled to a bosonic thermal reservoir. We describe the evolution of the density operator $\hat\chi(t)$ on the composite Hilbert space $\mathcal{H}_\mathrm{S}\otimes\mathcal{H}_\mathrm{A}$ (system $+$ spectator) in the Born-Markov and secular approximations by a Lindblad master equation. (Recall that the quantum state of the LZ qubit is then described by the density matrix $\hat{\rho}(t)={\rm Tr}_f\{\hat \chi(t)\}$ obtained by tracing out the degrees of freedom of the spectator field.) For a spectator implemented as a two-level system (auxiliary qubit) the dynamics read
\begin{equation}
    \frac{\mathrm{d}}{\mathrm{d}t}\hat{\chi}(t)
    = -i\big[\hat{H}(t),\hat{\chi}(t)\big]
      + \kappa\;\mathcal{L}_A\!\big[\hat\chi(t)\big],
    \label{eq:master_qubit}
\end{equation}
where $\hat H(t)$ denotes the full (system $+$ spectator) Hamiltonian introduced in the main text, $\kappa$ is the spectator damping (thermalization) rate, and the spectator Lindbladian $\mathcal{L}_A$ is written in terms of the usual dissipator $\mathcal{D}(o)[\cdot]$:
\begin{equation}
    \mathcal{L}_A[\chi] = (n+1)\,\mathcal{D}(\hat{o})[\chi] + n\,\mathcal{D}(\hat{o}^\dagger)[\chi],\qquad \mathcal{D}(\hat{o})[\chi] \equiv \hat{o}\,\chi\,\hat{o}^\dagger - \tfrac{1}{2}\{\hat{o}^\dagger\hat{o},\chi\}. \label{eq:lindblad_qubit}
\end{equation}
For the auxiliary qubit the jump operators are defined as $\hat{o}=\mathbb{I}_\mathrm{S}\otimes\hat\sigma_-$, $\hat{o}^\dagger=\mathbb{I}_\mathrm{S}\otimes\hat\sigma_+$ so that $\mathcal{D}(\hat{o})$ describes spontaneous emission of the quantum field, and $n=(\exp[\omega_/(k_B\Theta)]-1)^{-1}$ is the mean occupation number of the reservoir with frequency $\omega$ at temperature $\Theta$. In the zero-temperature limit $n\to 0$, Eq.~\eqref{eq:master_qubit} reduces to the simpler form
\begin{equation}
    \frac{\mathrm{d}}{\mathrm{d}t}\hat{\chi}
    = -i\big[\hat{H}(t),\hat{\chi}\big]
      + \kappa\!\left(\hat{o}\hat{\chi}\hat{o}^\dagger - \tfrac{1}{2}\{\hat{o}^\dagger\hat{o},\hat{\chi}\}\right),
    \label{eq:master_qubit_zeroT}
\end{equation}
which we implement in the numerics.

\begin{figure}[t]
    \centering
    \includegraphics[width=0.9\linewidth]{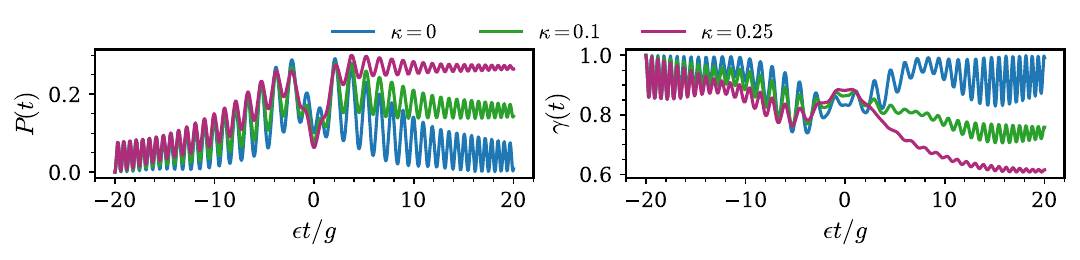}
    \includegraphics[width=0.9\linewidth]{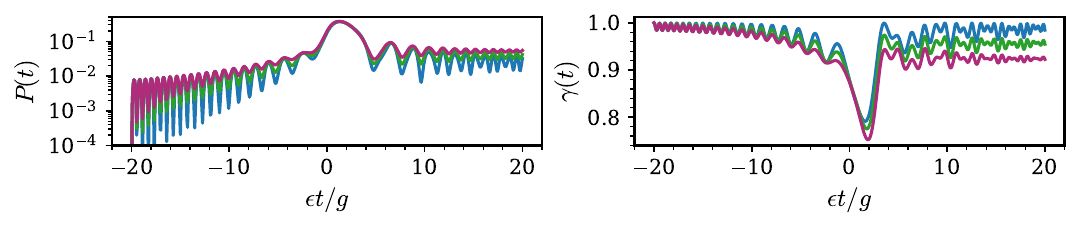}
    \caption{Dynamical response of the system to a lossy spectator, with dynamics described by Eq.~\eqref{eq:master_qubit_zeroT}, for three damping rates: $\kappa=0$ (blue), $\kappa=0.1$ (green), and  $\kappa=0.25$ (purple). Time is given in the dimensionless units $\epsilon t/g$ and the sweep runs from $t_i=-t_f$ to $t_f$ with $t_f=20g/\epsilon$. Top panel: the excitation probability $P(t)$ (left) and the purity $\gamma(t)$ (right) for the parameters $x_0=3g$, $\omega_c = x_0/4$. Bottom panel: the transition probability $P(t)$ (left) shown on a log scale to highlight low-infidelity features and the corresponding purity $\gamma(t)$ (right) for parameters  $x_0=g$, $\omega_c = 2.5\,x_0$, which lie in one of the low-infidelity ``islands’’. In all panels $\epsilon=2g^2$ is fixed. Weak spectator damping suppresses the oscillations in both $P(t)$ and $\gamma(t)$, but--- depending on the chosen $(x_0/g,\omega_c/x_0)$ point in parameter space---this smoothing can come at the cost of a reduced final fidelity (see Fig.~\ref{fig:dissipation} for the fidelity landscapes).}
    \label{fig:dissipation_cuts}
\end{figure}

The numeric results in Fig.~\ref{fig:dissipation} compare three cases: panel (a) reproduces the unitary (no-dissipation) data shown in Fig.~\ref{fig:sm2} of the Supplemental (and Fig.~4(a) of the main text) for reference; panels (b) and (c) show the same protocol when the spectator qubit is weakly lossy, with dynamics governed by Eq.~\eqref{eq:master_qubit_zeroT} and damping rates $\kappa=0.01$ and $\kappa=0.1$, respectively. The main qualitative effect of finite $\kappa$ is a smoothing of the fine, oscillatory structure in the fidelity landscape (i.e., reduced sensitivity to the precise values of coupling $x_0$, spectator frequency $\omega_c$ and final time $t_f$). However, this smoothing is accompanied by a systematic reduction in the maximum fidelity that can be achieved. Even though (stronger) damping does effectively suppress oscillations (see Fig.~\ref{fig:dissipation_cuts}), the associated loss of coherence and population from the joint system typically reduces the final fidelity and purity by an amount that outweighs any benefit of oscillation suppression for the finite annealing durations considered here. 

Interestingly, the effect of spectator loss is not uniform across parameter space: robustness to dissipation is markedly stronger when the protocol parameters lie inside one of the low-infidelity ``islands'' visible in the density plot of Fig.~\ref{fig:dissipation}.  As an illustrative example (bottom panel of Fig.~\ref{fig:dissipation_cuts}), weak-to-moderate dissipation visibly damps the oscillations while the infidelity remains suppressed by roughly one order of magnitude relative to the isolated Landau-Zener benchmark value of $P(t_f)\approx 0.45$. By contrast, points outside those islands are far more sensitive: the top panel of Fig.~\ref{fig:dissipation_cuts} exemplifies parameter values for which dissipation both damps oscillations and substantially degrades performance, so that no net fidelity gain remains. These observations imply a clear design principle for implementations in cavity platforms: to advantageously exploit a lossy spectator one should (i) choose protocol parameters that place the dynamics inside a robust low-infidelity island and (ii) ensure that the spectator damping rate $\kappa$ remains small compared to the relevant coherent energy scales of the coupled system. In practice, the trade-off between oscillation suppression and coherence loss must be evaluated quantitatively for each specific experimental realization and requires further investigation. Moreover, it would be interesting to study potential links between the influence of dissipation and spectral properties of the system.

In closing, the above discussion generalizes to the case where the spectator is a single bosonic mode (harmonic oscillator) with Hamiltonian $\hat{H}_f=\omega_c\hat a^\dagger\hat a$; see Sec.~\ref{sec:HMOspectator}. Modelling spectator loss with a thermal Lindblad master equation yields the same qualitative conclusion as for the lossy spectator qubit: a weakly lossy spectator tends to smooth the oscillatory, parameter-sensitive features of the unitary dynamics. This smoothing is generally accompanied by a reduction of final fidelity and purity, but there are regions in parameter space where a robustness to dissipation is exhibited. While qualitatively similar to the case of the qubit quantum field, we remark that the oscillator's unbounded Hilbert space does lead to important quantitative differences. 

\end{document}